\documentclass[journal]{IEEEtran}
\IEEEoverridecommandlockouts
\usepackage{cite}
\usepackage{amsmath,amssymb,amsfonts}
\usepackage{algorithmic}
\usepackage{graphicx}
\usepackage{textcomp}
\usepackage{xcolor}
\usepackage{cite}
\usepackage{amsmath,amssymb,amsfonts}
\usepackage{algorithm}
\usepackage{multirow}
\usepackage{siunitx}
\usepackage{algorithmic}
\usepackage{comment}
\usepackage{caption}
\usepackage{subcaption}
\usepackage{booktabs,
            makecell, 
            tabularx}
\usepackage{textcomp}
\usepackage{mathtools}
\def\BibTeX{{\rm B\kern-.05em{\sc i\kern-.025em b}\kern-.08em
    T\kern-.1667em\lower.7ex\hbox{E}\kern-.125emX}}
\ifCLASSINFOpdf
\else
\fi
\begin{document}
%
\title{At-Admission Prediction of Mortality and Pulmonary Embolism in  COVID-19 Patients Using Statistical and Machine Learning Methods: An International Cohort Study}
%
%

\author{Munib Mesinovic, Xin Ci Wong, Giri Shan Rajahram, Barbara Wanjiru Citarella, Kalaiarasu M. Peariasamy, Frank van Someren Gr\'eve, Piero Olliaro, Laura Merson, Lei Clifton, Christiana Kartsonaki, ISARIC Characterisation Group*
\thanks{M. Mesinovic is with the Department of Engineering Science, University of Oxford, Oxford, UK, e-mail: munib.mesinovic@jesus.ox.ax.uk}
\thanks{X. Wong is with the Digital Health Research and Innovation Unit, Institute for Clinical Research, National Institutes of Health (NIH), Malaysia}
\thanks{G. Rajahram is with the Queen Eliazabeth II Hospital, Ministry of Health, Malaysia}
\thanks{B. Citarella is with ISARIC, Pandemic Sciences Institute, University of Oxford, UK}
\thanks{K. Peariasamy is with the Institute for Clinical Research, National Institutes of Health (NIH), Ministry of Health Malaysia, Kuala Lumpur, Malaysia}
\thanks{F. Gr\'eve is with the Department of Medical Microbiology, Amsterdam University Medical Center, Amsterdam, The Netherlands}
\thanks{P. Olliaro is with ISARIC, Pandemic Sciences Institute, University of Oxford, UK}
\thanks{L. Merson is with ISARIC, Pandemic Sciences Institute, University of Oxford, UK}
\thanks{L. Clifton is with the Nuffield Department of Population Health, University of Oxford, Oxford, UK}
\thanks{C. Kartsonaki is with the MRC Population Health Research Unit, Clinical Trials Service Unit and Epidemiological Studies Unit, Nuffield Department of Population Health, University of Oxford, Oxford, UK}
\thanks{*listed in separate document}}

\maketitle

\begin{abstract}
By September, 2022, more than 600 million cases of SARS-CoV-2 infection have been reported globally, resulting in over 6.5 million deaths. COVID-19 mortality risk estimators are often, however, developed with small unrepresentative samples and with methodological limitations. It is highly important to develop predictive tools for pulmonary embolism (PE) in COVID-19 patients as one of the most severe preventable complications of COVID-19. Early recognition can help provide life-saving targeted anti-coagulation therapy right at admission. \\

\noindent Using a dataset of more than 800,000 COVID-19 patients from an international cohort, we propose a cost-sensitive gradient-boosted machine learning model that predicts occurrence of PE and death at admission. Logistic regression, Cox proportional hazards models, and Shapley values were used to identify key predictors for PE and death. \\

\noindent Our prediction model had a test AUROC of 75.9\% and 74.2\%, and sensitivities of 67.5\% and 72.7\% for PE and all-cause mortality respectively on a highly diverse and held-out test set. The PE prediction model was also evaluated on patients in UK and Spain separately with test results of 74.5\% AUROC, 63.5\% sensitivity and 78.9\% AUROC, 95.7\% sensitivity. Age, sex, region of admission, comorbidities (chronic cardiac and pulmonary disease, dementia, diabetes, hypertension, cancer, obesity, smoking), and symptoms (any, confusion, chest pain, fatigue, headache, fever, muscle or joint pain, shortness of breath) were the most important clinical predictors at admission. Age, overall presence of symptoms, shortness of breath, and hypertension were found to be key predictors for PE using our extreme gradient boosted model. \\

\noindent This analysis based on the, until now, largest global dataset for this set of problems can inform hospital prioritisation policy and guide long term clinical research and decision-making for COVID-19 patients globally. Our machine learning model developed from an international cohort can serve to better regulate hospital risk prioritisation of at-risk patients.
\end{abstract}

\begin{IEEEkeywords}
COVID-19, digital health, explainability, machine learning, mortality, pulmonary embolism, risk prediction, survival analysis
\end{IEEEkeywords}

\IEEEpeerreviewmaketitle

\section{Introduction}
\subsection{Clinical Background}
\noindent On the last day of 2019, the WHO received information about 44 cases of pneumonia-like disease in Wuhan city, China \cite{world2020novel}. By 5 September 2022, more than 600 million cases of SARS-CoV-2 infection had been reported across all continents, regions, and most countries, resulting in nearly 6.5 million deaths \cite{johns2020covid}. \\

\noindent COVID-19, the disease caused by infection with SARS-CoV-2, has a high mortality rate in hospitalised patients with deaths predominantly caused by respiratory failure \cite{yang2020clinical}. It continues to this day to be a challenging global pandemic with significant morbidity and mortality \cite{liao2020incidence}. As \cite{knight2021prospective} indicate, prognostic models that can predict outcomes among COVID-19 patients can be used to support clinical decision-making regarding hospital treatment and prioritisation. One such score is the 4C score that includes data about patient comorbidity, abnormal physiology, and inflammation using routinely measured data, bedside observations, and biochemistry tests \cite{jones2021external}. While in most cases COVID-19 is a mild illness, those at highest risk of death and severe complications usually are hospitalised some time after onset \cite{tabata2020clinical}.\\

\noindent Pulmonary embolism (PE) is among the most severe and preventable complications of COVID-19 characterized by increased D-dimer levels and high thrombosis risk that has been repeatedly reported across different countries \cite{susen2020prevention}. Studies suggest PE incidence rates above 15\% in the ICU for COVID-19 patients and early recognition of its risk factors can help in identifying urgent treatment with anticoagulation therapy to those most in clinical need \cite{liao2020incidence, whiteley2022risk}. Recent international studies additionally suggest COVID-19 as a key risk factor for pulmonary embolism both in the short- and long-term \cite{whiteley2022risk, katsoularis2022risks}. Existing PE prediction models are limited in part because they were developed for non-COVID-19 patients and traditional risk factors for PE may not be as predictive. If risk models can be developed for assessing occurrence of PE in COVID-19 patients across different countries, that can be an important step forward in preventing this serious complication of COVID-19, especially given the current epidemiological situation \cite{whiteley2022risk}. \\

\noindent As for risk factors that contribute the most to the occurrence of mortality and pulmonary embolism in COVID-19 patients, age has been established as the dominant predictor of mortality \cite{marcos2021development}. Furthermore, studies have described other risk factors of COVID mortality such as cardiovascular disease, chronic respiratory disease, diabetes, hypertension, smoking, and obesity \cite{venturini2021classification}. \\

\subsection{Technical Background}
\noindent Machine learning has been applied to different COVID-19 related questions. Large amounts of patient data are being generated during the COVID-19 pandemic which can be useful for predictive modelling. Using machine learning with large amounts of complex patient data could generate accurate and patient-specific predictions and assist clinicians. \\ 

Previous research includes \cite{zhou2020clinical} exploring in-hospital mortality with logistic regression on just 191 patients and \cite{xie2020development} have followed with multi-center validation with 299 patients for internal training and 145 patients for external validaton. \cite{alaa2020retrospective} have looked at regression-based predictions of all-cause mortality with hospital admission time as a predictor and using hazard models yet their results have also been limited due to a smaller dataset restricting generalisability. All of these studies have used a combination of demographics, comorbidities, symptoms, laboratory tests, and self-reported onset times. \\

\noindent In this study, we investigated how pulmonary embolism and all-cause mortality vary across subgroups of a large and international cohort. We also show how predictive certain clinical factors gathered from patients with COVID-19 can be to the respective outcomes. In studies looking at predicting thromboembolism more broadly, a defining limitation for impactful and generalisable application of machine learning methods has been a small patient sample and a lack of systematic comparison of algorithms \cite{van2020predicting}. Applying a diverse set of methods to one of the largest and most diverse datasets on hospitalised patients with COVID-19 can help find the best mechanism for risk prioritisation of patients in a timely way and may help reduce mortality and risk of PE in those with COVID-19. \\

\section{Methodology}
\subsection{Data}
\noindent In this work, we use data of COVID-19 patients from The International Severe Acute Respiratory and Emerging Infection Consortium (ISARIC), a repository that standardises and secures data on COVID-19 assembled from a global cohort over 2 years of the pandemic as of January 2022. It includes so far data on over 800,000 patients from 53 countries. These data capture the global experience of the first 2 years of the pandemic \cite{akhvlediani2020isaric}. The clinical characterisation protocol underwent ethical review by the World Health Organization Ethics Review Committee and ethics approval was obtained for each participating country and site according to local requirements. \\

\noindent The study population consisted of all patients with either clinically diagnosed or laboratory confirmed COVID-19 admitted to the participating hospitals. The aim of the recruiting sites was to use a consecutive sample. \\

\noindent The dataset contains 800,459 patients and 182 variables. The mean age of patients was 56.4 (20.9), 48.6\% were male, and the majority of cases were from South Africa (54.0\%) and the United Kingdom (34.1\%). 65.3\% of patients were discharged alive and 20.4\% died. We grouped countries with fewer than 60 individuals into a single category. Out of all patients, 5,450 (0.7\%) experienced a pulmonary embolism, 73 experienced thromboembolism, and 143 experienced Deep Vein Thrombosis. We define our outcome of interest as the main pulmonary embolism (PE) diagnosis for subsequent analysis. 4,653 (82.1\%) of the PE cases were recorded in the United Kingdom (UK) and Spain which based on our knowledge makes our study the largest study of its kind for PE to date. Due to similar data collection patterns and recording, we used data from these two countries only for PE modelling as they contain the vast majority of reported PE cases. \\

\subsection{Data Preprocessing}
\noindent Since treatment information does not have reliable timestamps for most patients, the following variables were used in the analysis for PE: demographics (including age, sex, country), comorbidities (hypertension, diabetes, smoking etc.), and symptoms (coughing, fever, fatigue etc.). The presence of diagnosis during domination by the alpha variant was also included (after December 2020) due to its possible association with incidence of PE. In our modelling of PE, we used data from patients from the UK and Spain only and did not use laboratory measurements or imputation methods. 269,373 patients and 45 variables remained for PE, and 734,282 patients and 55 variables for death. Age was grouped into 5 categories (0-20, 20-40, 40-60, 60-80, 80-120) with the distributions seen in Figures \ref{fig:age_death} and \ref{fig:age_SpainUK} below. The \textit{symptomatic} variable represents any symptoms reported for a patient. For number of days from admission to event (death), we removed outliers of more than 200 days and those in the negatives, thereby removing 1,342 patients. \\

\begin{figure}[H]
    \centering
    \includegraphics[width=0.9\linewidth]{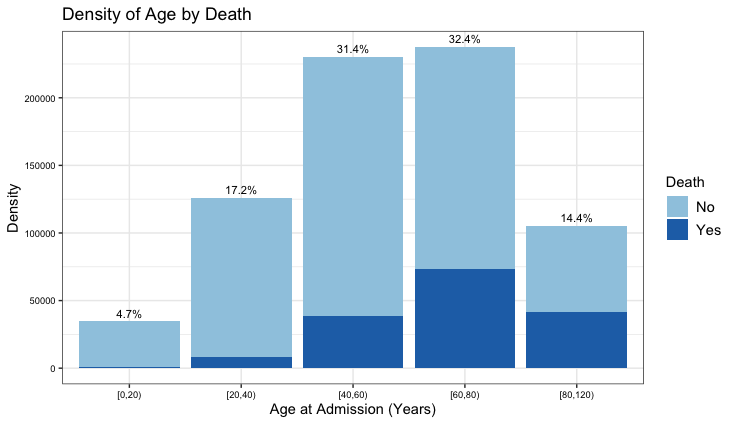}
    \caption{Age Distribution for All Patients Stratified by Death Outcome}
    \label{fig:age_death}
\end{figure}

\begin{figure}[H]
    \centering
    \includegraphics[width=0.9\linewidth]{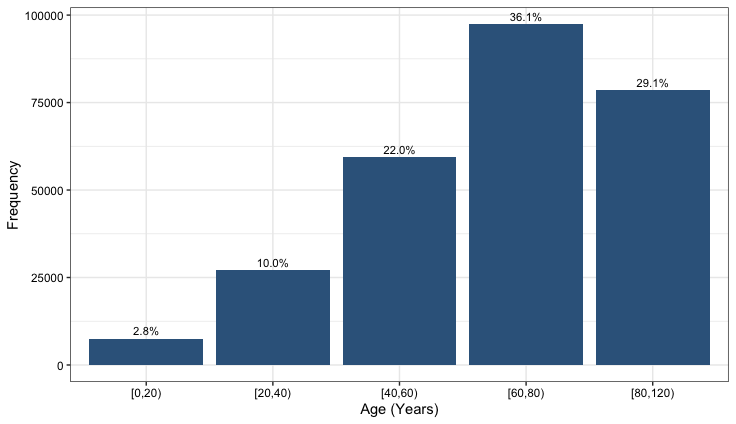}
    \caption{Age Distribution for UK and Spain Patients}
    \label{fig:age_SpainUK}
\end{figure}

\begin{figure}[H]
    \centering
    \includegraphics[width=0.9\linewidth]{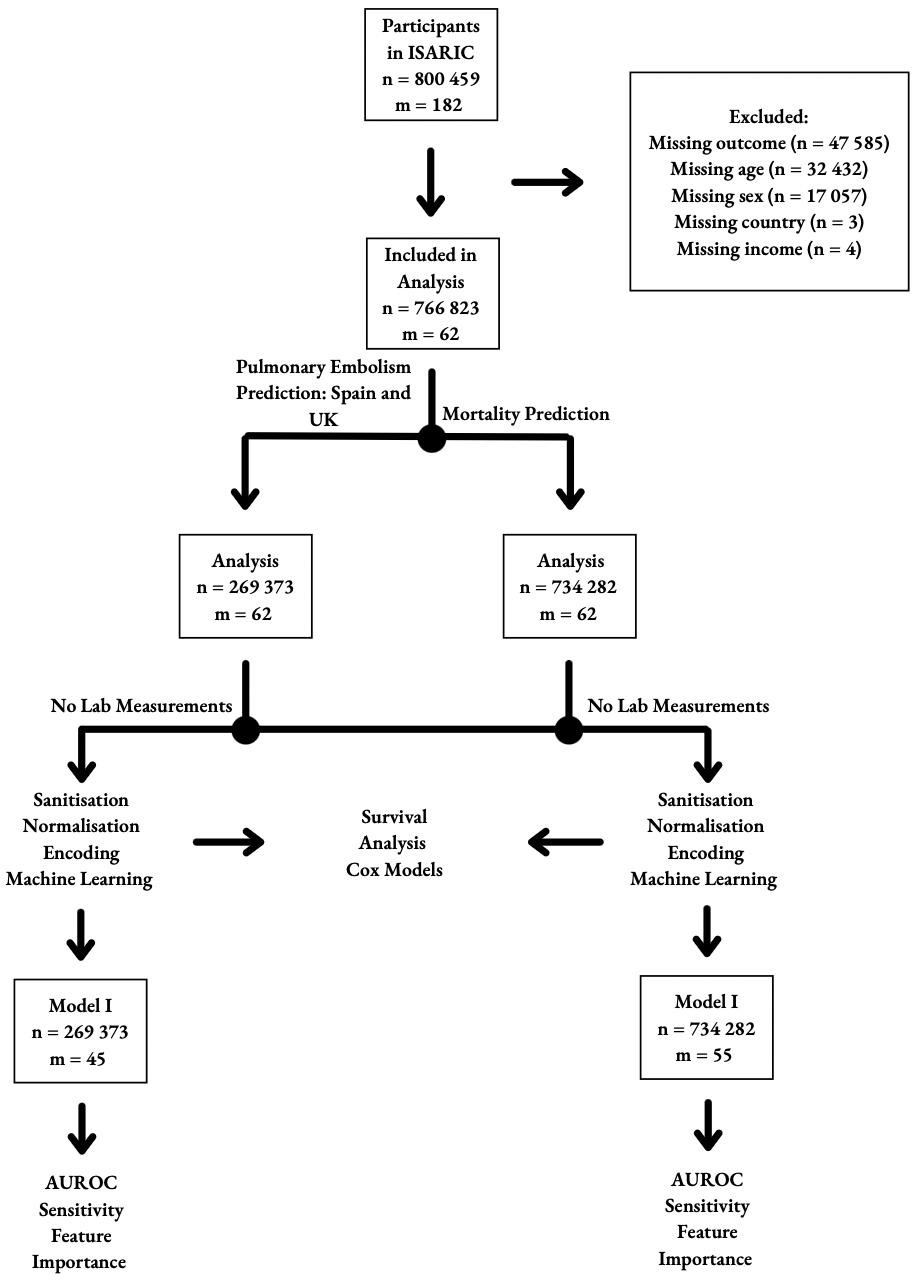}
    \caption{Flowchart of Framework with Machine Learning Model to Predict The Risk of PE And Mortality At Admission}
    \label{fig:flowchart}
\end{figure}

\noindent Prior to processing the data, for PE prediction, we held out 3 test sets of 20\% of the total dataset for independent testing, one of which would only include patients from Spain, one only including patients from the UK, and another including both. For mortality prediction testing, we held out 20\% of the total dataset sample. A workflow diagram for data processing and system design is shown in Figure \ref{fig:flowchart}. \\

\noindent Stratified Kaplan--Meier curves by age, sex, and region of admission were plotted using Cox Proportional Hazards Models while machine learning methods were applied for prediction of PE or death. \\

\subsection{Baseline and Machine Learning Methods}
\noindent We investigated several prediction methods for PE occurrence and death, including logistic regression, Linear Discriminant Analysis (LDA), naive Bayes classifier, random forests, ADABoosting algorithms, and the high-performing extreme gradient boost machine (XGBoost) \cite{kumari2017machine, chowdhury2021early}. Previous studies looking at tree-based algorithms such as XGBoost have highlighted its capacity to learn the correlations between covariates well when it comes to mortality prediction in COVID-19 patients while also being somewhat interpretable \cite{baqui2021comparing}. We applied all of these methods for the purposes of a systematic comparison using cross-validation, several hold-out test sets stratified across countries and regions, and evaluated with multiple metrics. A list of methods applied can be seen in Table \ref{tab:models} with details in Supplementary. \\

\begin{table}
\caption{\textbf{Machine learning methods investigated}}
\small
\centering
\begin{tabular}{@{}ll@{}}
\toprule
    \textbf{Models} & \textbf{Brief Description} \\
    \midrule
Logistic Regression & generalised linear model \\
\addlinespace[0.05cm]
Linear Discriminant Analysis & normal distribution, linear \\
\addlinespace[0.05cm]
Naive Bayes & independence, probabilistic \\
\addlinespace[0.05cm]
Random Forest & decision tree ensemble \\
\addlinespace[0.05cm]
XGBoost & gradient-boosted decision trees \\
\addlinespace[0.05cm]
Ensemble & AdaBoost models ensemble \\
\addlinespace[0.05cm]
Ensemble with XGBoost & XGBoost models ensemble \\
\bottomrule                          
\end{tabular}
\label{tab:models}
\end{table}

\noindent As is often the case in disease prediction, there is class imbalance with about 1.7\% of UK and Spain patients having been diagnosed with PE and around 20.4\% having died in the case population. To address this, we use other metrics mentioned above in the evaluation of our models besides accuracy as it does not capture the true predictive performance of our models and we rely more on sensitivity and the F1 score. We also use a different threshold for prediction after probability estimation instead of the default 0.5 to achieve cost-sensitivity, and we apply random undersampling at a minority:majority ratio of 1:4 as has been highlighted in other work \cite{ling2008cost, lemaitre2017imbalanced}. We evaluate these methods both separately and in combination to investigate the best approach for this set of prediction problems. \\

\noindent To address imbalance in predictions, we applied either undersampling, thresholding, or both. As for death, due to a much softer imbalance, undersampling was not necessary. We also added class-weighting to our XGBoost model using inverse proportions and compared it with the other methods to address class imbalance. All confusion matrices and parameter details for each model can be found in the Supplementary. \\

\noindent Furthermore, we build an ensemble that combines AdaBoosted decision trees with robust undersampling using different subsets for resampled training so as to address the imbalance and compare this cost-sensitive model with our best performing model and add further confidence in its ability to generalise in an imbalanced scenario \cite{liu2009easyensemble}. We extend the ensemble learning methods by using our own XGBoost model in the ensemble structure instead of the standard AdaBoosted decision trees. The number of trees was a tunable hyperparameter listed in the Supplementary (Tables IV-VI). We compare our cost-sensitive class-weighted XGBoost machine learning model with these resampling ensembles to show improved performance without the need of introducing bias like in the case of resampling while maintainting interpretability. \\

\subsection{Model Validation and Evaluation}
\noindent We proceed to tune our machine learning models and validate them using stratified 5-fold cross-validation with Bayesian optimisation. We repeated the optimisation procedure for 50 iterations after which we evaluated the model on the independent test set with the following metrics: AUROC, Accuracy, Weighted F1 Score, and Sensitivity. The details can be found in the Supplementary. \\

\noindent While existing studies referenced in the Introduction mention some approaches to feature importance estimation for COVID-19 mortality and outcome prediction as well as for other problems, rarely does one find several interpretability methods compared and contrasted in one scenario. We implemented both tree-based F-score interpretability methods as well as Shapley values analysis, logistic regression and Cox regression, and hope to draw interesting conclusions from each and their comparisons \cite{lundberg2020local, ibrahim2020explainable}. A full explanation for the Shapley values method and its details can be found in the Supplementary materials. \\

\section{Role of the funding source}
\noindent The funder had no role in study design, data collection, data analysis, data interpretation, writing of the report, and decision to submit the paper for publication. \\

\section{Results}
\noindent Variable distributions can be seen in Tables \ref{tab:Characteristics1}, \ref{tab:Characteristics2}, \ref{tab:Characteristics1Death}, and \ref{tab:Characteristics2Death}. A detailed collection of figures for variable distribution across age groups can be found in the Supplementary. \\

\begin{table*}[t]
    \small
    \centering
    \caption{\textbf{Baseline Characteristics Stratified by Occurrence of PE (median and IQR used for lab measurements). PE here also includes positive cases of deep vein thrombosis and thromboembolism}}
    \begin{tabularx}{\textwidth}{l
                                    r
                                    r
                                    r
                                    r
                                    r
                                    r
                                 }
         \toprule
        & \multicolumn{2}{c}{PE (N = 5,656)} && \multicolumn{2}{c}{Non-PE}  \\
        \cmidrule(r){2-3} \cmidrule(l){5-6}
         \bf{Characteristic}  &  \multicolumn{1}{c}{\bf{Missing (\%)}} & \bf{Mean (SD)/count (\%)} &&  \multicolumn{1}{c}{\bf{\bf{Missing (\%)}}} & \bf{Mean (SD)/count (\%)}\\
            \midrule
          Age & 0.8 & 62.6 (15.6) && 4.1 & 56.4 (20.9)\\
          \addlinespace[0.05cm]
          Sex (male) & 0.1 & 3,733 (66.0) && 2.1 & 385,038 (48.4)\\
          \addlinespace[0.05cm]
          Alpha Variant (post) & - & 3,733 (66.0) && 2.1 & 385,038 (48.4)\\
          \addlinespace[0.05cm]
          \textbf{Ethnicity} &  &   &&  & \\
          \addlinespace[0.05cm]
          \hspace{0.5cm} White & - & 309 (5.5)  && - & 12,030 (1.5) \\
          \addlinespace[0.05cm]
          \hspace{0.5cm} South Asian & - & 32 (0.6)  && - & 9,224 (1.2) \\
          \addlinespace[0.05cm]
          \hspace{0.5cm} Malay & - & 0 (0.0)  && - & 3,812 (0.5) \\
          \addlinespace[0.05cm]
          \hspace{0.5cm} Latin American & - & 24 (0.4)  && - & 2,719 (0.3) \\
          \addlinespace[0.05cm]
          \hspace{0.5cm} Other & - & 5,171 (91.4)  && - & 757,656 (95.3) \\
          \addlinespace[0.05cm]
          \textbf{Country} &  &   &&  & \\
          \addlinespace[0.05cm]
          \hspace{0.5cm} South Africa & - & 0 (0.0) && - & 432,596 (54.4) \\
          \addlinespace[0.05cm]
          \hspace{0.5cm} United Kingdom & - & 4,076 (72.1)  && - & 269,073 (33.9) \\
          \addlinespace[0.05cm]
          \hspace{0.5cm} Spain & - & 577 (10.2)  && - & 14,764 (1.9) \\
          \addlinespace[0.05cm]
          \hspace{0.5cm} Norway & - & 15 (0.3)  && - & 7,448 (0.9) \\
          \addlinespace[0.05cm]
          \hspace{0.5cm} Other & - & 988 (17.4)  && - & 71,853 (9.0) \\
          \addlinespace[0.05cm]
          \textbf{Country Income} &  &   &&  & \\
          \addlinespace[0.05cm]
          \hspace{0.5cm} High & - & 5,513 (97.5)  && - & 323,889 (40.8) \\
          \addlinespace[0.05cm]
          \hspace{0.5cm} Upper Middle & - & 72 (1.3)  && - & 449,782 (56.6) \\
          \addlinespace[0.05cm]
          \hspace{0.5cm} Lower Middle & - & 71 (1.2)  && - & 20,210 (2.5) \\
          \addlinespace[0.05cm]
          \hspace{0.5cm} Low & - & 0 (0.0)  && - & 918 (0.1) \\
          \addlinespace[0.05cm]
          \textbf{Region} &  &   &&  & \\
          \addlinespace[0.05cm]
          \hspace{0.5cm} Sub-saharan Africa & - & 0 (0.0)  && - & 433,522 (54.5) \\
          \addlinespace[0.05cm]
          \hspace{0.5cm} Europe and Central Asia & - & 5,264 (93.1)  && - & 314,416 (39.6) \\
          \addlinespace[0.05cm]
          \hspace{0.5cm} South Asia & - & 49 (0.9)  && - & 17,413 (2.2) \\
          \addlinespace[0.05cm]
          \hspace{0.5cm} East Asia & - & 39 (0.7)  && - & 10,421 (1.3) \\
          \addlinespace[0.05cm]
          \hspace{0.5cm} North America & - & 195 (3.4)  && - & 9,666 (1.2) \\
          \addlinespace[0.05cm]
          \hspace{0.5cm} Other & - & 107 (1.9)  && - & 9,498 (1.2) \\
          \addlinespace[0.05cm]
          \textbf{Comorbidities} &  &   &&  & \\
          \addlinespace[0.05cm]
          \hspace{0.5cm} AIDS/HIV & 15.5 & 19 (0.3)  && 28.3 & 27,895 (3.5) \\
          \addlinespace[0.05cm]
          \hspace{0.5cm} Asthma & 10.6 & 644 (11.4)  && 26.4 & 51,714 (6.5) \\
          \addlinespace[0.05cm]
          \hspace{0.5cm} Chronic Cardiac Disease & 8.3 & 1,002 (17.7)  && 26.2 & 80,348 (10.1) \\
          \addlinespace[0.05cm]
          \hspace{0.5cm} Chronic Haematological & 11.2 & 171 (3.0)  && 64.7 & 10,275 (1.3) \\
          \addlinespace[0.05cm]
          \hspace{0.5cm} Chronic Kidney Disease & 8.8 & 516 (9.1)  && 27.0 & 46,054 (5.8) \\
          \addlinespace[0.05cm]
          \hspace{0.5cm} Chronic Neurological & 9.6 & 369 (6.5)  && 63.8 & 28,648 (3.6) \\
          \addlinespace[0.05cm]
          \hspace{0.5cm} Chronic Pulmonary & 7.9 & 707 (12.5)  && 26.5 & 52,006 (6.5) \\
          \addlinespace[0.05cm]
          \hspace{0.5cm} Dementia & 9.8 & 214 (3.8)  && 64.5 & 27,374 (3.4) \\
          \addlinespace[0.05cm]
          \hspace{0.5cm} Diabetes & 10.4 & 1,196 (21.1)  && 24.5 & 152,728 (19.2) \\
          \hspace{0.5cm} Hypertension & 10.3 & 2,219 (39.2)  && 27.1 & 226,285 (28.5) \\
          \addlinespace[0.05cm]
          \hspace{0.5cm} Liver Disease & 7.3 & 169 (3.0)  && 62.3 & 8,960 (1.1) \\
          \addlinespace[0.05cm]
          \hspace{0.5cm} Malignant Neoplasm & 8.7 & 508 (9.0)  && 27.1 & 27,811 (3.5) \\
          \addlinespace[0.05cm]
          \hspace{0.5cm} Malnutrition & 15.1 & 59 (1.0)  && 67.3 & 5,582 (0.7) \\
          \addlinespace[0.05cm]
          \hspace{0.5cm} Obesity & 16.2 & 1,167 (20.6)  && 57.0 & 53,227 (6.7) \\
          \addlinespace[0.05cm]
          \hspace{0.5cm} Rheumatologic & 9.9 & 485 (8.6)  && 64.9 & 27,954 (3.5) \\
          \addlinespace[0.05cm]
          \hspace{0.5cm} Smoking & 45.5 & 1,317 (23.3)  && 71.5 & 74,205 (9.3) \\
          \addlinespace[0.05cm]
            \bottomrule
    \end{tabularx}
    \label{tab:Characteristics1}
\end{table*}

\begin{table*}[t]
    \small
    \centering
    \caption{\textbf{Baseline Characteristics Stratified by Occurrence of PE (continued). PE here also includes positive cases of deep vein thrombosis and thromboembolism}}
    \begin{tabularx}{\textwidth}{l
                                    r
                                    r
                                    r
                                    r
                                    r
                                    r
                                 }
         \toprule
        & \multicolumn{2}{c}{PE (N = 5,656)} && \multicolumn{2}{c}{Non-PE}  \\
        \cmidrule(r){2-3} \cmidrule(l){5-6}
         \bf{Characteristic}  &  \multicolumn{1}{c}{\bf{Missing (\%)}} & \bf{Mean (SD)/count (\%)} &&  \multicolumn{1}{c}{\bf{\bf{Missing (\%)}}} & \bf{Mean (SD)/count (\%)}\\
            \midrule
          \textbf{Symptoms} & &  &&  & \\
          \addlinespace[0.05cm]
          \hspace{0.5cm} Symptomatic & 2.3 & 5,400 (95.5)  && 63.4 & 276,645 (34.8) \\
          \addlinespace[0.05cm]
          \hspace{0.5cm} Abdominal Pain & 19.1 & 310 (5.5)  && 70.6 & 21,510 (2.7) \\
          \addlinespace[0.05cm]
          \hspace{0.5cm} Confusion & 14.4 & 624 (11.0)  && 70.5 & 45,559 (5.7) \\
          \addlinespace[0.05cm]
          \hspace{0.5cm} Bleeding & 19.0 & 95 (1.7)  && 70.8 & 4,307 (0.5) \\
          \addlinespace[0.05cm]
          \hspace{0.5cm} Chest Pain & 17.0 & 1,058 (18.7)  && 70.4 & 32,811 (4.1) \\
          \addlinespace[0.05cm]
          \hspace{0.5cm} Conjunctivitis & 23.2 & 15 (0.3)  && 72.0 & 1,012 (0.1) \\
          \addlinespace[0.05cm]
          \hspace{0.5cm} Cough & 10.7 & 3,434 (60.7)  && 67.7 & 153,548 (19.3) \\
          \addlinespace[0.05cm]
          \hspace{0.5cm} Diarrhoea & 14.1 & 886 (15.7)  && 69.9 & 39,718 (5.0) \\
          \addlinespace[0.05cm]
          \hspace{0.5cm} Ear Pain & 43.4 & 7 (0.1)  && 77.0 & 765 (0.1) \\
          \addlinespace[0.05cm]
          \hspace{0.5cm} Fatigue & 18.5 & 2,134 (37.7)  && 70.9 & 93,637 (11.8) \\
          \addlinespace[0.05cm]
          \hspace{0.5cm} Headache & 20.0 & 510 (9.0)  && 71.8 & 26,815 (3.4) \\
          \addlinespace[0.05cm]
          \hspace{0.5cm} Fever & 10.6 & 3,029 (53.6)  && 67.8 & 146,248 (18.4) \\
          \addlinespace[0.05cm]
          \hspace{0.5cm} Lost Sense of Smell & 24.7 & 403 (7.1)  && 77.4 & 12,736 (1.6) \\
          \addlinespace[0.05cm]
          \hspace{0.5cm} Lost Sense of Taste & 28.3 & 438 (7.7)  && 77.9 & 14,767 (1.9) \\
          \addlinespace[0.05cm]
          \hspace{0.5cm} Lymphadenopathy & 23.7 & 33 (0.6)  && 72.9 & 1,264 (0.2) \\
          \addlinespace[0.05cm]
          \hspace{0.5cm} Muscle/Joint Pain & 20.5 & 903 (16.0)  && 71.8 & 40,627 (5.1) \\
          \addlinespace[0.05cm]
          \hspace{0.5cm} Runny Nose & 26.0 & 111 (2.0)  && 73.0 & 8,244 (1.0) \\
          \addlinespace[0.05cm]
          \hspace{0.5cm} Seizures & 18.1 & 19 (0.3)  && 71.3 & 2,764 (0.3) \\
          \addlinespace[0.05cm]
          \hspace{0.5cm} Severe Dehydration & 64.9 & 245 (4.3) && 85.3 & 13,732 (1.7) \\
          \addlinespace[0.05cm]
          \hspace{0.5cm} Shortness of Breath & 6.8 & 4,205 (74.3)  && 67.5 & 156,078 (19.6) \\
          \addlinespace[0.05cm]
          \hspace{0.5cm} Skin Rash & 21.5 & 76 (1.3)  && 71.5 & 5,771 (0.7) \\
          \addlinespace[0.05cm]
          \hspace{0.5cm} Sore Throat & 25.8 & 225 (4.0)  && 72.9 & 16,500 (2.1) \\
          \addlinespace[0.05cm]
          \hspace{0.5cm} Vomiting & 17.3 & 736 (13.0)  && 69.9 & 43,149 (5.4) \\
          \addlinespace[0.05cm]
          \hspace{0.5cm} Wheezing & 22.7 & 278 (4.9)  && 71.7 & 14,474 (1.8) \\
          \addlinespace[0.05cm]
          \textbf{Lab Measurements} & &  &&  & \\
          \addlinespace[0.05cm]
          \hspace{0.5cm} D-dimer ($\mu$g/mL) & 92.0 & 1.1 (0.5, 2.5)  && 98.6 & 0.7 (0.4, 1.3) \\
          \addlinespace[0.05cm]
          \hspace{0.5cm} ALT (IU/L) & 64.1 & 36.0 (22.0, 60.0)  && 86.0 & 27.0 (17.0, 45.0) \\
          \addlinespace[0.05cm]
          \hspace{0.5cm} Bilirubin ($\mu$mol/L) & 70.5 & 11.0 (8.0, 15.6)  && 85.7 & 9.0 (7.0, 14.0) \\
          \addlinespace[0.05cm]
          \hspace{0.5cm} CRP (mg/L) & 35.7 & 115.0 (59.0, 191.8)  && 79.7 & 74.9 (29.0, 143.0) \\
          \addlinespace[0.05cm]
          \hspace{0.5cm} Lymphocytes ($[10^3$]/$\mu$L) & 33.2 & 0.9 (0.6, 1.3)  && 78.4 & 0.9 (0.6, 1.3) \\
          \addlinespace[0.05cm]
          \hspace{0.5cm} Neutrophils ($[10^9$]/L) & 33.2 & 7.1 (4.8, 10.1)  && 78.4 & 5.5 (3.8, 8.2) \\
          \addlinespace[0.05cm]
          \hspace{0.5cm} Platelets ($[10^9$]/L) & 57.7 & 13.0 (11.5, 14.7)  && 88.0 & 12.8 (11.2, 14.3) \\
          \addlinespace[0.05cm]
          \hspace{0.5cm} Blood Urea Nitrogen (mmol/L) & 44.1 & 6.5 (4.7, 9.7)  && 79.8 & 6.4 (4.5, 10.0) \\
          \addlinespace[0.05cm]
          \hspace{0.5cm} White Blood Cells ($[10^9$]/L) & 30.0 & 8.8 (6.2, 12.1)  && 77.0 & 7.2 (5.3, 10.1) \\
          \addlinespace[0.05cm]
          \textbf{Vital Signs} & &  &&  & \\
          \addlinespace[0.05cm]
          \hspace{0.5cm} Diastolic BP (mmHg) & 6.9 & 75.5 (14.8)  && 64.7 & 74.8 (15.2) \\
          \addlinespace[0.05cm]
          \hspace{0.5cm} Systolic BP (mmHg) & 6.8 & 130.2 (23.2) && 64.7 & 130.3 (24.5) \\
          \addlinespace[0.05cm]
          \hspace{0.5cm} Heart Rate (bpm) & 7.4 & 96.3 (21.1)  && 65.4 & 92.0 (21.5) \\
          \addlinespace[0.05cm]
          \hspace{0.5cm} Oxygen Saturation (\%) & 6.8 & 90.7 (11.3)  && 64.7 & 93.4 (9.3) \\
          \addlinespace[0.05cm]
          \hspace{0.5cm} Respiratory Rate (brpm) & 10.1 & 25.1 (8.0)  && 65.7 & 22.8 (7.0) \\
          \addlinespace[0.05cm]
          \hspace{0.5cm} Temperature ($^{\circ}$C) & 7.1 & 37.2 (1.1)  && 64.2 & 37.2 (1.0) \\
          \addlinespace[0.05cm]
          \textbf{Outcome} & &  &&  &  \\
          \addlinespace[0.05cm]
          \hspace{0.5cm} Discharge & - & 3,492 (61.7)  && - & 519,423 (65.4) \\
          \addlinespace[0.05cm]
          \hspace{0.5cm} Death & - & 1,297 (22.9)  && - & 162,091 (20.4) \\
          \addlinespace[0.05cm]
          \hspace{0.5cm} Other & - & 531 (15.4)  && - & 54,930 (14.2) \\
          \addlinespace[0.05cm]
            \bottomrule
    \end{tabularx}
    \label{tab:Characteristics2}
\end{table*}

\begin{table*}[t]
    \small
    \centering
    \caption{\textbf{Baseline Characteristics Stratified by Death (median and IQR used for lab measurements)}}
    \begin{tabularx}{\textwidth}{l
                                    r
                                    r
                                    r
                                    r
                                    r
                                    r
                                 }
         \toprule
        & \multicolumn{2}{c}{Death (N = 163,388)} && \multicolumn{2}{c}{No Death}  \\
        \cmidrule(r){2-3} \cmidrule(l){5-6}
         \bf{Characteristic}  &  \multicolumn{1}{c}{\bf{Missing (\%)}} & \bf{Mean (SD)/count (\%)} &&  \multicolumn{1}{c}{\bf{\bf{Missing (\%)}}} & \bf{Mean (SD)/count (\%)}\\
            \midrule
          Age & 0.8 & 67.5 (16.1) && 2.8 & 53.2 (21.0)\\
          \addlinespace[0.05cm]
          Sex (male) & 0.2 & 87,679 (53.7) && 0.5 & 282,993 (48.0)\\
          \addlinespace[0.05cm]
          \textbf{Country} &  &   &&  & \\
          \addlinespace[0.05cm]
          \hspace{0.5cm} South Africa & - & 96,965 (59.3) && - & 332,012 (56.3) \\
          \addlinespace[0.05cm]
          \hspace{0.5cm} United Kingdom & - & 54,540 (33.4)  && - & 193,780 (32.9) \\
          \addlinespace[0.05cm]
          \hspace{0.5cm} Spain & - & 123 (0.1)  && - & 7,486 (1.3) \\
          \addlinespace[0.05cm]
          \hspace{0.5cm} Norway & - & 41 ($<$0.1)  && - & 7,216 (1.2) \\
          \addlinespace[0.05cm]
          \hspace{0.5cm} Other & - & 11,719 (7.2)  && - & 48,992 (8.3) \\
          \addlinespace[0.05cm]
          \textbf{Country Income} &  &   &&  & \\
          \addlinespace[0.05cm]
          \hspace{0.5cm} High & - & 59,531 (36.4)  && - & 231,101 (39.2) \\
          \addlinespace[0.05cm]
          \hspace{0.5cm} Upper Middle & - & 98,064 (60.0)  && - & 345,013 (58.5) \\
          \addlinespace[0.05cm]
          \hspace{0.5cm} Lower Middle & - & 5,725 (3.5)  && - & 12,654 (2.1) \\
          \addlinespace[0.05cm]
          \hspace{0.5cm} Low & - & 68 ($<$0.1)  && - & 717 (0.1) \\
          \addlinespace[0.05cm]
          \textbf{Region} &  &   &&  & \\
          \addlinespace[0.05cm]
          \hspace{0.5cm} Sub-saharan Africa & - & 97,033 (59.4)  && - & 332,730 (56.4) \\
          \addlinespace[0.05cm]
          \hspace{0.5cm} Europe and Central Asia & - & 57,129 (35.0)  && - & 222,268 (37.7) \\
          \addlinespace[0.05cm]
          \hspace{0.5cm} South Asia & - & 5,232 (3.2)  && - & 11,491 (1.9) \\
          \addlinespace[0.05cm]
          \hspace{0.5cm} East Asia & - & 678 (0.4)  && - & 8,322 (1.4) \\
          \addlinespace[0.05cm]
          \hspace{0.5cm} North America & - & 2,099 (1.3)  && - & 6,776 (1.1) \\
          \addlinespace[0.05cm]
          \hspace{0.5cm} Other & - & 1,237 (0.8)  && - & 7,999 (1.4) \\
          \addlinespace[0.05cm]
          \textbf{Comorbidities} &  &   &&  & \\
          \addlinespace[0.05cm]
          \hspace{0.5cm} AIDS/HIV & 27.3 & 6,203 (3.8)  && 26.2 & 20,828 (3.5) \\
          \addlinespace[0.05cm]
          \hspace{0.5cm} Asthma & 25.2 & 10,425 (6.4) && 24.6 & 40,402 (6.9) \\
          \addlinespace[0.05cm]
          \hspace{0.5cm} Chronic Cardiac Disease & 24.4 & 26,400 (16.2)  && 24.6 & 52,337 (8.9) \\
          \addlinespace[0.05cm]
          \hspace{0.5cm} Chronic Haematological & 65.1 & 2,901 (1.8)  && 65.8 & 7,095 (1.2) \\
          \addlinespace[0.05cm]
          \hspace{0.5cm} Chronic Kidney Disease & 25.6 & 16,693 (10.2)  && 25.3 & 28,559 (4.8) \\
          \addlinespace[0.05cm]
          \hspace{0.5cm} Chronic Neurological & 64.1 & 8,071 (4.9)  && 64.7 & 19,981 (3.4) \\
          \addlinespace[0.05cm]
          \hspace{0.5cm} Chronic Pulmonary & 25.0 & 16,224 (9.9) && 24.7 & 34,869 (5.9) \\
          \addlinespace[0.05cm]
          \hspace{0.5cm} Dementia & 65.1 & 10,261 (6.3)  && 65.5 & 16,740 (2.8) \\
          \addlinespace[0.05cm]
          \hspace{0.5cm} Diabetes & 21.3 & 45,593 (27.9)  && 23.1 & 104,810 (17.8) \\
          \hspace{0.5cm} Hypertension & 24.7 & 64,098 (39.2)  && 25.4 & 157,991 (26.8) \\
          \addlinespace[0.05cm]
          \hspace{0.5cm} Immunosuppression & 79.6 & 1,364 (0.8)  && 81.9 & 3,728 (0.6) \\
          \addlinespace[0.05cm]
          \hspace{0.5cm} Liver Disease & 62.4 & 2,262 (1.4)  && 63.3 & 6,269 (1.1) \\
          \addlinespace[0.05cm]
          \hspace{0.5cm} Malignant Neoplasm & 25.3 & 9,036 (5.5)  && 25.5 & 17,719 (3.0) \\
          \addlinespace[0.05cm]
          \hspace{0.5cm} Malnutrition & 67.0 & 1,717 (1.1)  && 67.1 & 3,692 (0.6) \\
          \addlinespace[0.05cm]
          \hspace{0.5cm} Obesity & 55.1 & 11,176 (6.8)  && 56.3 & 39,153 (6.6) \\
          \addlinespace[0.05cm]
          \hspace{0.5cm} Rheumatologic & 65.2 & 7,333 (4.5) && 65.9 & 20,359 (3.5) \\
          \addlinespace[0.05cm]
          \hspace{0.5cm} Smoking & 72.5 & 17,655 (10.8)  && 71.5 & 53,764 (9.1) \\
          \addlinespace[0.05cm]
          \hspace{0.5cm} Tuberculosis & 55.5 & 2,725 (1.7)  && 55.6 & 8,688 (1.5) \\
          \addlinespace[0.05cm]
            \bottomrule
    \end{tabularx}
    \label{tab:Characteristics1Death}
\end{table*}

\begin{table*}[t]
    \small
    \centering
    \caption{\textbf{Baseline Characteristics Stratified by Death (continued). PE here also includes positive cases of deep vein thrombosis and thromboembolism}}
    \begin{tabularx}{\textwidth}{l
                                    r
                                    r
                                    r
                                    r
                                    r
                                    r
                                 }
         \toprule
        & \multicolumn{2}{c}{Death (N = 163,388)} && \multicolumn{2}{c}{No Death}  \\
        \cmidrule(r){2-3} \cmidrule(l){5-6}
         \bf{Characteristic}  &  \multicolumn{1}{c}{\bf{Missing (\%)}} & \bf{Mean (SD)/count (\%)} &&  \multicolumn{1}{c}{\bf{\bf{Missing (\%)}}} & \bf{Mean (SD)/count (\%)}\\
            \midrule
          \textbf{Symptoms} & &  &&  & \\
          \addlinespace[0.05cm]
          \hspace{0.5cm} Symptomatic & 63.9 & 57,578 (35.2)  && 61.7 & 211,379 (35.9) \\
          \addlinespace[0.05cm]
          \hspace{0.5cm} Abdominal Pain & 71.4 & 3,452 (2.1)  && 69.1 & 17,537 (3.0) \\
          \addlinespace[0.05cm]
          \hspace{0.5cm} Confusion & 69.7 & 16,267 (10.0)  && 69.3 & 29,022 (4.9) \\
          \addlinespace[0.05cm]
          \hspace{0.5cm} Bleeding & 71.3 & 1,099 (0.7)  && 69.3 & 3,202 (0.5) \\
          \addlinespace[0.05cm]
          \hspace{0.5cm} Chest Pain & 71.4 & 4,554 (2.8)  && 68.7 & 27,581 (4.7) \\
          \addlinespace[0.05cm]
          \hspace{0.5cm} Conjunctivitis & 72.8 & 220 (0.1)  && 70.6 & 792 (0.1) \\
          \addlinespace[0.05cm]
          \hspace{0.5cm} Cough & 68.2 & 31,110 (19.0) && 66.0 & 117,745 (20.0) \\
          \addlinespace[0.05cm]
          \hspace{0.5cm} Diarrhoea & 70.5 & 6,860 (4.2)  && 68.3 & 31,776 (5.4) \\
          \addlinespace[0.05cm]
          \hspace{0.5cm} Ear Pain & 76.7 & 100 (0.1)  && 75.4 & 629 (0.1) \\
          \addlinespace[0.05cm]
          \hspace{0.5cm} Fatigue & 71.7 & 19,510 (11.9)  && 69.4 & 70,064 (11.9) \\
          \addlinespace[0.05cm]
          \hspace{0.5cm} Headache & 73.4 & 2,469 (1.5) && 70.1 & 22,228 (3.8) \\
          \addlinespace[0.05cm]
          \hspace{0.5cm} Fever & 68.0 & 29,252 (17.9)  && 66.3 & 111,358 (18.9) \\
          \addlinespace[0.05cm]
          \hspace{0.5cm} Lost Sense of Smell & 79.6 & 1,131 (0.7)  && 75.6 & 10,998 (1.9) \\
          \addlinespace[0.05cm]
          \hspace{0.5cm} Lost Sense of Taste & 80.2 & 1,671 (1.0)  && 76.2 & 12,648 (2.1) \\
          \addlinespace[0.05cm]
          \hspace{0.5cm} Lymphadenopathy & 72.8 & 311 (0.2)  && 71.8 & 843 (0.1) \\
          \addlinespace[0.05cm]
          \hspace{0.5cm} Muscle/Joint Pain & 73.5 & 5,706 (3.5)  && 70.1 & 33,181 (5.6) \\
          \addlinespace[0.05cm]
          \hspace{0.5cm} Runny Nose & 74.1 & 865 (0.5)  && 71.5 & 6,058 (1.0) \\
          \addlinespace[0.05cm]
          \hspace{0.5cm} Seizures & 71.2 & 573 (0.4)  && 70.1 & 2,109 (0.4) \\
          \addlinespace[0.05cm]
          \hspace{0.5cm} Severe Dehydration & 85.5 & 4,400 (2.7) && 84.2 & 9,369 (1.6) \\
          \addlinespace[0.05cm]
          \hspace{0.5cm} Shortness of Breath & 67.6 & 37,074 (22.7)  && 65.9 & 116,392 (19.7) \\
          \addlinespace[0.05cm]
          \hspace{0.5cm} Skin Rash & 72.1 & 1,593 (1.0)  && 70.1 & 4,028 (0.7) \\
          \addlinespace[0.05cm]
          \hspace{0.5cm} Sore Throat & 74.0 & 2,115 (1.3)  && 71.4 & 12,826 (2.2) \\
          \addlinespace[0.05cm]
          \hspace{0.5cm} Vomiting & 70.6 & 6,506 (4.0)  && 68.4 & 35,578 (6.0) \\
          \addlinespace[0.05cm]
          \hspace{0.5cm} Wheezing & 72.3 & 4,077 (2.5)  && 70.2 & 9,779 (1.7) \\
          \addlinespace[0.05cm]
          \textbf{Lab Measurements} & &  &&  & \\
          \addlinespace[0.05cm]
          \hspace{0.5cm} D-dimer ($\mu$g/mL) & 99.1 & 1.1 (0.5, 2.5)  && 98.6 & 0.7 (0.4, 1.3) \\
          \addlinespace[0.05cm]
          \hspace{0.5cm} ALT (IU/L) & 85.3 & 36.0 (22.0, 60.0)  && 85.2 & 27.0 (17.0, 45.0) \\
          \addlinespace[0.05cm]
          \hspace{0.5cm} Bilirubin ($\mu$mol/L) & 84.3 & 11.0 (8.0, 15.6)  && 85.2 & 9.0 (7.0, 14.0) \\
          \addlinespace[0.05cm]
          \hspace{0.5cm} CRP (mg/L) & 78.9 & 115.0 (59.0, 191.8)  && 78.3 & 74.9 (29.0, 143.0) \\
          \addlinespace[0.05cm]
          \hspace{0.5cm} Lymphocytes ($[10^3]$$\mu$L) & 78.3 & 0.9 (0.6, 1.3)  && 76.7 & 0.9 (0.6, 1.3) \\
          \addlinespace[0.05cm]
          \hspace{0.5cm} Neutrophils ($[10^9$]/L) & 78.3 & 7.1 (4.8, 10.1)  && 76.7 & 5.5 (3.8, 8.2) \\
          \addlinespace[0.05cm]
          \hspace{0.5cm} Platelets ($[10^9$]/L) & 86.7 & 13.0 (11.5, 14.7)  && 87.3 & 12.8 (11.2, 14.3) \\
          \addlinespace[0.05cm]
          \hspace{0.5cm} Blood Urea Nitrogen (mmol/L) & 78.9 & 6.5 (4.7, 9.7)  && 78.5 & 6.4 (4.5, 10.0) \\
          \addlinespace[0.05cm]
          \hspace{0.5cm} White Blood Cells ($[10^9$]/L) & 76.6 & 8.8 (6.2, 12.1)  && 75.3 & 7.2 (5.3, 10.1) \\
          \addlinespace[0.05cm]
          \textbf{Vital Signs} & &  &&  & \\
          \addlinespace[0.05cm]
          \hspace{0.5cm} Diastolic BP (mmHg) & 63.1 & 72.9 (16.3)  && 64.3 & 75.5 (14.8) \\
          \addlinespace[0.05cm]
          \hspace{0.5cm} Systolic BP (mmHg) & 63.0 & 130.0 (26.3) && 64.2 & 131.0 (24.0) \\
          \addlinespace[0.05cm]
          \hspace{0.5cm} Heart Rate (bpm) & 63.2 & 92.5 (22.3)  && 65.3 & 91.8 (21.2) \\
          \addlinespace[0.05cm]
          \hspace{0.5cm} Oxygen Saturation ($\%$) & 63.3 & 91.3 (10.2)  && 64.3 & 93.8 (9.2) \\
          \addlinespace[0.05cm]
          \hspace{0.5cm} Respiratory Rate (brpm) & 63.4 & 24.4 (7.7)  && 64.7 & 22.3 (6.6) \\
          \addlinespace[0.05cm]
          \hspace{0.5cm} Temperature ($^{\circ}$C) & 63.1 & 37.2 (1.1)  && 63.9 & 37.2 (1.0) \\
          \addlinespace[0.05cm]
          \textbf{PE} & &  &&  &  \\
          \addlinespace[0.05cm]
          \hspace{0.5cm} Yes & 99.2 & 1,365 (0.8)  && 99.3 & 4,291 (0.7) \\
          \addlinespace[0.05cm]
            \bottomrule
    \end{tabularx}
    \label{tab:Characteristics2Death}
\end{table*}

\begin{figure*}[t]
    \centering
    \includegraphics[width=1.0\linewidth]{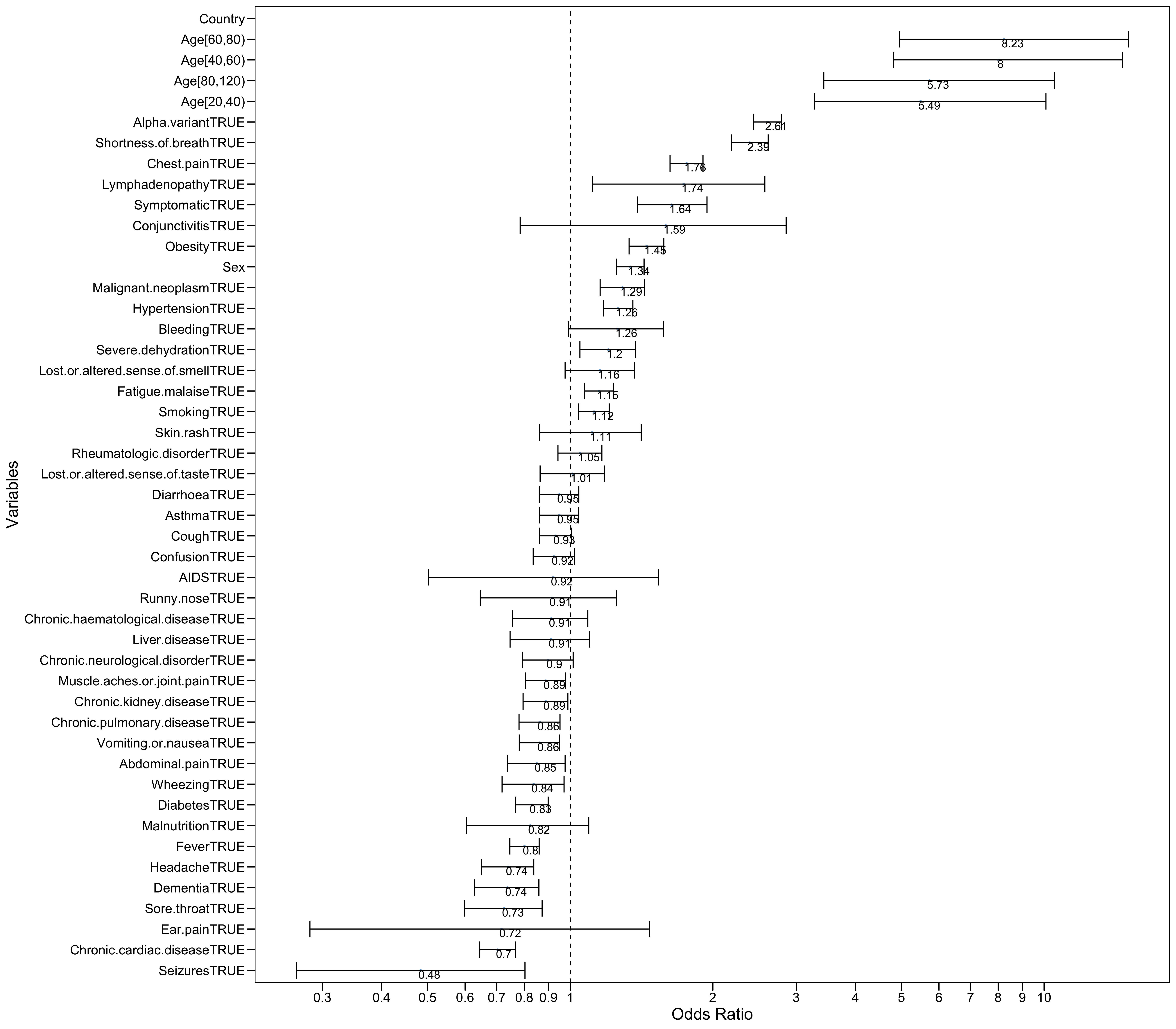}
    \caption{Logistic Regression Odds Ratios for PE Outcome}
    \label{fig:oddsPE}
\end{figure*}

\begin{figure*}[t]
    \centering
    \includegraphics[width=1.0\linewidth]{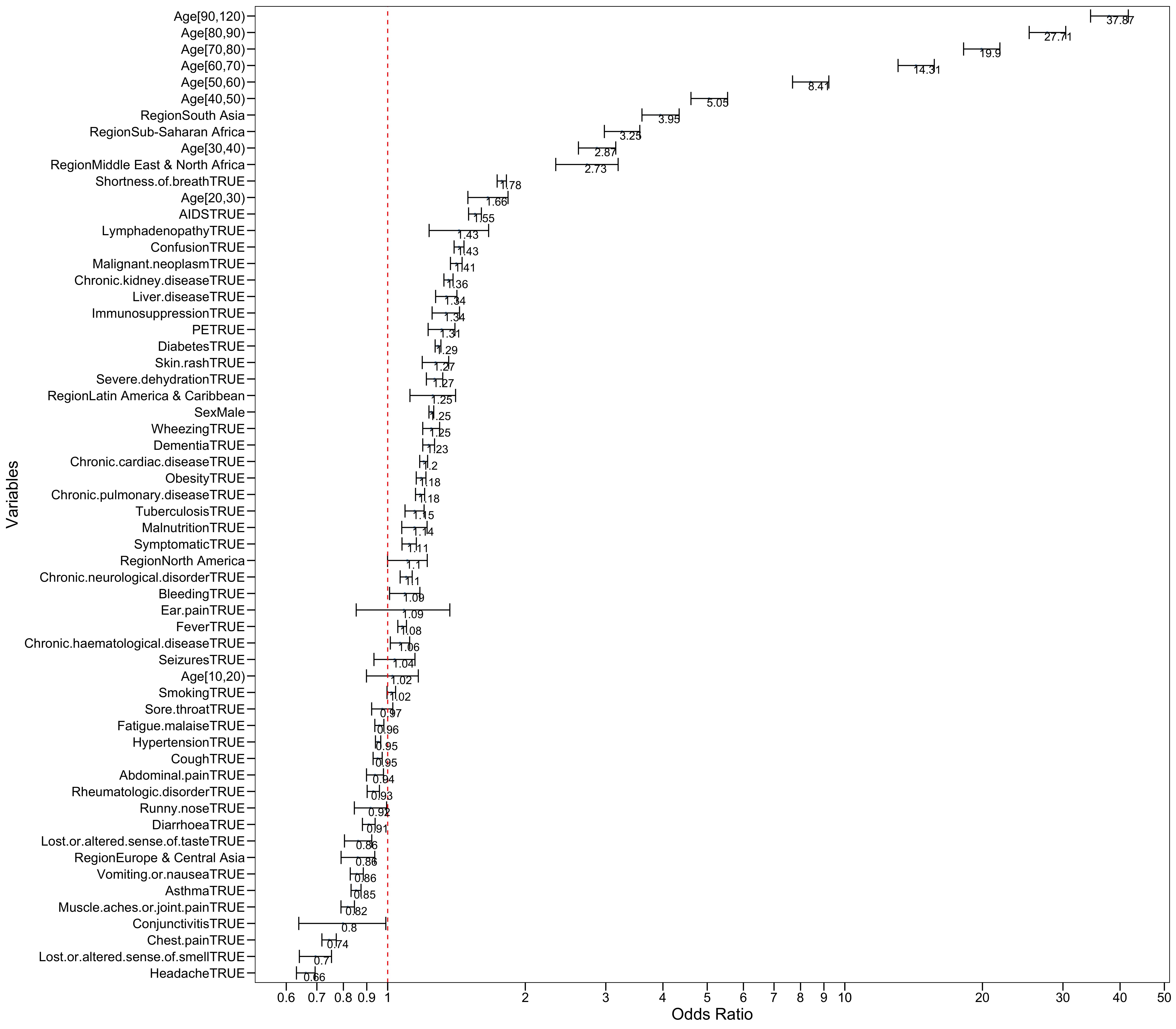}
    \caption{Logistic Regression Odds Ratios for Death Outcome}
    \label{fig:oddsDeath}
\end{figure*}

\noindent Several variables were highly correlated with PE and death (Supplementary Figures 3 and 4, Tables II and III). Multivariable logistic regression shows high association of country, age, alpha variant, and certain symptoms with PE and death (Figures \ref{fig:oddsPE} and \ref{fig:oddsDeath}). Tables with $p$-values are included in Tables \ref{tab:Odds1}, \ref{tab:Odds2}, and \ref{tab:Odds3}. \\

\begin{table*}[t]
    \centering
    \small
    \caption{\textbf{Multivariable Logistic Regression Odds Ratios of Features with 95\% Confidence Intervals (only Spain and UK patients included for PE)}}
    \begin{tabularx}{0.7\linewidth}{l
                                    r
                                    r
                                    r
                                    r
                                    r
                                    r
                                 }
         \toprule
        & \multicolumn{2}{c}{PE} && \multicolumn{2}{c}{Death}  \\
        \cmidrule(r){2-3} \cmidrule(l){5-6}
         \bf{Feature}  &  \multicolumn{1}{c}{\bf{OR (95\% CI)}} & \bf{P Value} &&  \multicolumn{1}{c}{\bf{\bf{OR (95\% CI)}}} & \bf{P Value}\\
            \midrule
          Age &  & $<$0.005 &&  & $<$0.005\\
          \addlinespace[0.05cm]
          \hspace{0.5cm} $<$20 & 1.0 & $<$0.005 && 1.0 & $<$0.005 \\
          \addlinespace[0.05cm]
          \hspace{0.5cm} 20-40 & 5.5 (3.3, 10.1) & $<$0.005  && 2.4 (2.3, 2.6) & $<$0.005\\
          \addlinespace[0.05cm]
          \hspace{0.5cm} 40-60 & 8.0 (4.8, 14.6) & $<$0.005  && 6.8 (6.4, 7.3) & $<$0.005 \\
          \addlinespace[0.05cm]
          \hspace{0.5cm} 60-80 & 8.2 (5.0, 15.1) & $<$0.005  && 16.1 (15.1, 17.2) & $<$0.005 \\
          \addlinespace[0.05cm]
          \hspace{0.5cm} $>$80 & 5.7 (3.4, 10.5) & $<$0.005  && 27.8 (26.0, 29.7) & $<$0.005 \\
          \addlinespace[0.05cm]
          Sex (male) & 1.3 (1.3, 1.4) & $<$0.005 && 1.2 (1.2, 1.3) & $<$0.005\\
          \addlinespace[0.05cm]
          \textbf{Region} &  &   &&  & $<$0.005\\
          \addlinespace[0.05cm]
          \hspace{0.5cm} Sub-saharan Africa & - & - && 3.3 (3.0, 3.6) & $<$0.005 \\
          \addlinespace[0.05cm]
          \hspace{0.5cm} Europe and Central Asia & - & -  && 0.9 (0.8, 1.0) & 0.052 \\
          \addlinespace[0.05cm]
          \hspace{0.5cm} South Asia & - & -  && 4.0 (3.7, 4.4) & $<$0.005 \\
          \addlinespace[0.05cm]
          \hspace{0.5cm} East Asia & - & -  && 1.0 & -\\
          \addlinespace[0.05cm]
          \hspace{0.5cm} North America & - & -  && 1.2 (1.1, 1.3) & $<$0.005 \\
          \addlinespace[0.05cm]
          \hspace{0.5cm} MENA & - & -  && 2.8 (2.4, 3.3) & $<$0.005 \\
          \addlinespace[0.05cm]
          Alpha Variant & 2.6 (2.4, 2.8) &  $<$0.005 &&  & \\
          \addlinespace[0.05cm]
          \textbf{Comorbidities} &  &   &&  & \\
          \addlinespace[0.05cm]
          \hspace{0.5cm} AIDS/HIV & 0.9 (0.5, 1.5) & 0.553 && 1.5 (1.4, 1.5) & $<$0.005 \\
          \addlinespace[0.05cm]
          \hspace{0.5cm} Asthma & 0.9 (0.9, 1.0) & 0.055 && 0.8 (0.8, 0.9) & $<$0.005\\
          \addlinespace[0.05cm]
          \hspace{0.5cm} Chronic Cardiac Disease & 0.7 (0.6, 0.8) & $<$0.005 && 1.2 (1.2, 1.2) & $<$0.005 \\
          \addlinespace[0.05cm]
          \hspace{0.5cm} Chronic Haematological & 0.9 (0.8, 1.1) & 0.320 && 1.1 (1.0, 1.1) & 0.016 \\
          \addlinespace[0.05cm]
          \hspace{0.5cm} Chronic Kidney Disease & 0.9 (0.8, 1.0) & $<$0.005 && 1.4 (1.3, 1.4) & $<$0.005 \\
          \addlinespace[0.05cm]
          \hspace{0.5cm} Chronic Neurological & 0.9 (0.8, 1.0) & 0.087 && 1.1 (1.1, 1.1) & $<$0.005 \\
          \addlinespace[0.05cm]
          \hspace{0.5cm} Chronic Pulmonary & 0.9 (0.8, 0.9) & $<$0.005 && 1.2 (1.2, 1.2) & $<$0.005 \\
          \addlinespace[0.05cm]
          \hspace{0.5cm} Dementia & 0.7 (0.6, 0.8) & $<$0.005 && 1.3 (1.2, 1.3) & $<$0.005 \\
            \bottomrule
    \end{tabularx}
    \label{tab:Odds1}
\end{table*}

\begin{table*}[t]
    \small
    \centering
    \caption{\textbf{Multivariable Logistic Regression Odds Ratios of Features with 95\% Confidence Intervals (only Spain and UK patients included for PE)}}
    \begin{tabularx}{0.7\linewidth}{l
                                    r
                                    r
                                    r
                                    r
                                    r
                                    r
                                 }
         \toprule
        & \multicolumn{2}{c}{PE} && \multicolumn{2}{c}{Death}  \\
        \cmidrule(r){2-3} \cmidrule(l){5-6}
         \bf{Feature}  &  \multicolumn{1}{c}{\bf{OR (95\% CI)}} & \bf{P Value} &&  \multicolumn{1}{c}{\bf{\bf{OR (95\% CI)}}} & \bf{P Value}\\
            \midrule
          \hspace{0.5cm} Diabetes & 0.8 (0.8, 0.9) & $<$0.005 && 1.3 (1.3, 1.3) & $<$0.005 \\
          \addlinespace[0.05cm]
          \hspace{0.5cm} Hypertension & 1.3 (1.2, 1.4) & $<$0.005 && 1.0 (1.0, 1.0) & $<$0.005 \\
          \addlinespace[0.05cm]
          \hspace{0.5cm} Liver Disease & 0.9 (0.7, 1.1) & 0.200 && 1.3 (1.2, 1.4) & $<$0.005 \\
          \addlinespace[0.05cm]
          \hspace{0.5cm} Malignant Neoplasm & 1.3 (1.2, 1.4) & $<$0.005 && 1.4 (1.4, 1.5) & $<$0.005 \\
          \addlinespace[0.05cm]
          \hspace{0.5cm} Malnutrition & 0.8 (0.6, 1.1) & 0.200 && 1.2 (1.1, 1.2) & $<$0.005 \\
          \addlinespace[0.05cm]
          \hspace{0.5cm} Obesity & 1.4 (1.3, 1.6) & $<$0.005 && 1.2 (1.1, 1.2) & $<$0.005 \\
          \addlinespace[0.05cm]
          \hspace{0.5cm} Rheumatologic & 1.1 (0.9, 1.2) & 0.137 && 0.9 (0.9, 1.0) & $<$0.005 \\
          \addlinespace[0.05cm]
          \hspace{0.5cm} Smoking & 1.1 (1.0, 1.2) & $<$0.005 && 1.0 (1.0, 1.0) & 0.126\\
          \addlinespace[0.05cm]
          \textbf{Symptoms} & &  &&  & \\
          \addlinespace[0.05cm]
          \hspace{0.5cm} Symptomatic & 1.6 (1.4, 1.9) & $<$0.005 && 1.1 (1.1, 1.1) & $<$0.005 \\
          \addlinespace[0.05cm]
          \hspace{0.5cm} Abdominal Pain & 0.8 (0.7, 0.9) & $<$0.005 && 0.9 (0.9, 1.0) & $<$0.005 \\
          \addlinespace[0.05cm]
          \hspace{0.5cm} Confusion & 0.9 (0.8, 1.0) & 0.009 && 1.5 (1.4, 1.5) & $<$0.005 \\
          \addlinespace[0.05cm]
          \hspace{0.5cm} Bleeding & 1.3 (1.0, 1.6) & 0.011 && 1.1 (1.0, 1.2) & 0.025 \\
          \addlinespace[0.05cm]
          \hspace{0.5cm} Chest Pain & 1.8 (1.6, 1.9) & $<$0.005 && 0.7 (0.7, 0.8) & $<$0.005 \\
          \addlinespace[0.05cm]
          \hspace{0.5cm} Conjunctivitis & 1.6 (0.8, 2.9) & 0.260 && 0.8 (0.6, 1.0) & 0.050 \\
          \addlinespace[0.05cm]
          \hspace{0.5cm} Cough & 0.9 (0.9, 1.0) & 0.391 && 0.9 (0.9, 1.0) & $<$0.005\\
          \addlinespace[0.05cm]
          \hspace{0.5cm} Diarrhoea & 0.9 (0.9, 1.1) & 0.568 && 0.9 (0.9, 0.9) & $<$0.005\\
          \addlinespace[0.05cm]
          \hspace{0.5cm} Ear Pain & 0.7 (0.3, 1.5) & 0.245 && 1.1 (0.9, 1.4) & 0.495 \\
          \addlinespace[0.05cm]
          \hspace{0.5cm} Fatigue & 1.1 (1.1, 1.2) & $<$0.005 && 1.0 (0.9, 1.0) & $<$0.005 \\
          \addlinespace[0.05cm]
          \hspace{0.5cm} Headache & 0.7 (0.7, 0.8) & $<$0.005 && 0.7 (0.6, 0.7) & $<$0.005 \\
          \addlinespace[0.05cm]
          \hspace{0.5cm} Fever & 0.8 (0.7, 0.9) & $<$0.005 && 1.1 (1.0, 1.1) & $<$0.005 \\
          \addlinespace[0.05cm]
          \hspace{0.5cm} Lost Sense of Smell & 1.2 (1.0, 1.4) & 0.012 && 0.7 (0.6, 0.7) & $<$0.005 \\
          \addlinespace[0.05cm]
          \hspace{0.5cm} Lost Sense of Taste & 1.0 (0.9, 1.2) & 0.490 && 0.9 (0.8, 0.9) & $<$0.005 \\
          \addlinespace[0.05cm]
          \hspace{0.5cm} Lymphadenopathy & 1.7 (1.1, 2.6) & $<$0.005 && 1.4 (1.2, 1.7) & $<$0.005 \\
          \addlinespace[0.05cm]
          \hspace{0.5cm} Muscle/Joint Pain & 0.9 (0.6, 1.0) & $<$0.005 && 0.8 (0.8, 0.8) & $<$0.005 \\
          \addlinespace[0.05cm]
          \hspace{0.5cm} Runny Nose & 0.9 (0.6, 1.3) & 0.832 && 0.9 (0.8, 1.0) & 0.042 \\
          \addlinespace[0.05cm]
          \hspace{0.5cm} Seizures & 0.5 (0.3, 0.8) & $<$0.005 && 1.0 (0.9, 1.1) & 0.784 \\
          \addlinespace[0.05cm]
          \hspace{0.5cm} Severe Dehydration & 1.2 (1.0, 1.4) & $<$0.005 && 1.3 (1.2, 1.3) & $<$0.005\\
            \bottomrule
    \end{tabularx}
    \label{tab:Odds2}
\end{table*}

\begin{table*}[t]
    \small
    \centering
    \caption{\textbf{Mulivariable Logistic Regression Odds Ratios of Features with 95\% Confidence Intervals (only Spain and UK patients included for PE)}}
    \begin{tabularx}{0.7\linewidth}{l
                                    r
                                    r
                                    r
                                    r
                                    r
                                    r
                                 }
         \toprule
        & \multicolumn{2}{c}{PE} && \multicolumn{2}{c}{Death}  \\
        \cmidrule(r){2-3} \cmidrule(l){5-6}
         \bf{Feature}  &  \multicolumn{1}{c}{\bf{OR (95\% CI)}} & \bf{P Value} &&  \multicolumn{1}{c}{\bf{\bf{OR (95\% CI)}}} & \bf{P Value}\\
            \midrule
          \addlinespace[0.05cm]
          \hspace{0.5cm} Shortness of Breath & 2.4 (2.2, 2.6) & $<$0.005 && 1.8 (1.7, 1.8) & $<$0.005 \\
          \addlinespace[0.05cm]
          \hspace{0.5cm} Skin Rash & 1.1 (0.9, 1.4) & 0.842 && 1.3 (1.2, 1.4) & $<$0.005 \\
          \addlinespace[0.05cm]
          \hspace{0.5cm} Sore Throat & 0.7 (0.6, 0.8) & $<$0.005 && 1.0 (0.9, 1.0) & 0.181 \\
          \addlinespace[0.05cm]
          \hspace{0.5cm} Vomiting & 0.9 (0.8, 1.0) & $<$0.005 && 0.9 (0.8, 0.9) & $<$0.005 \\
          \addlinespace[0.05cm]
          \hspace{0.5cm} Wheezing & 0.8 (0.7, 1.0) & $<$0.005  && 1.3 (1.2, 1.3) & $<$0.005 \\
          \addlinespace[0.05cm]
          PE & - & - && 1.3 (1.2, 1.4) & $<$0.005 \\
            \bottomrule
    \end{tabularx}
    \label{tab:Odds3}
\end{table*}

\begin{figure}[H]
    \centering
    \includegraphics[width=1.0\linewidth]{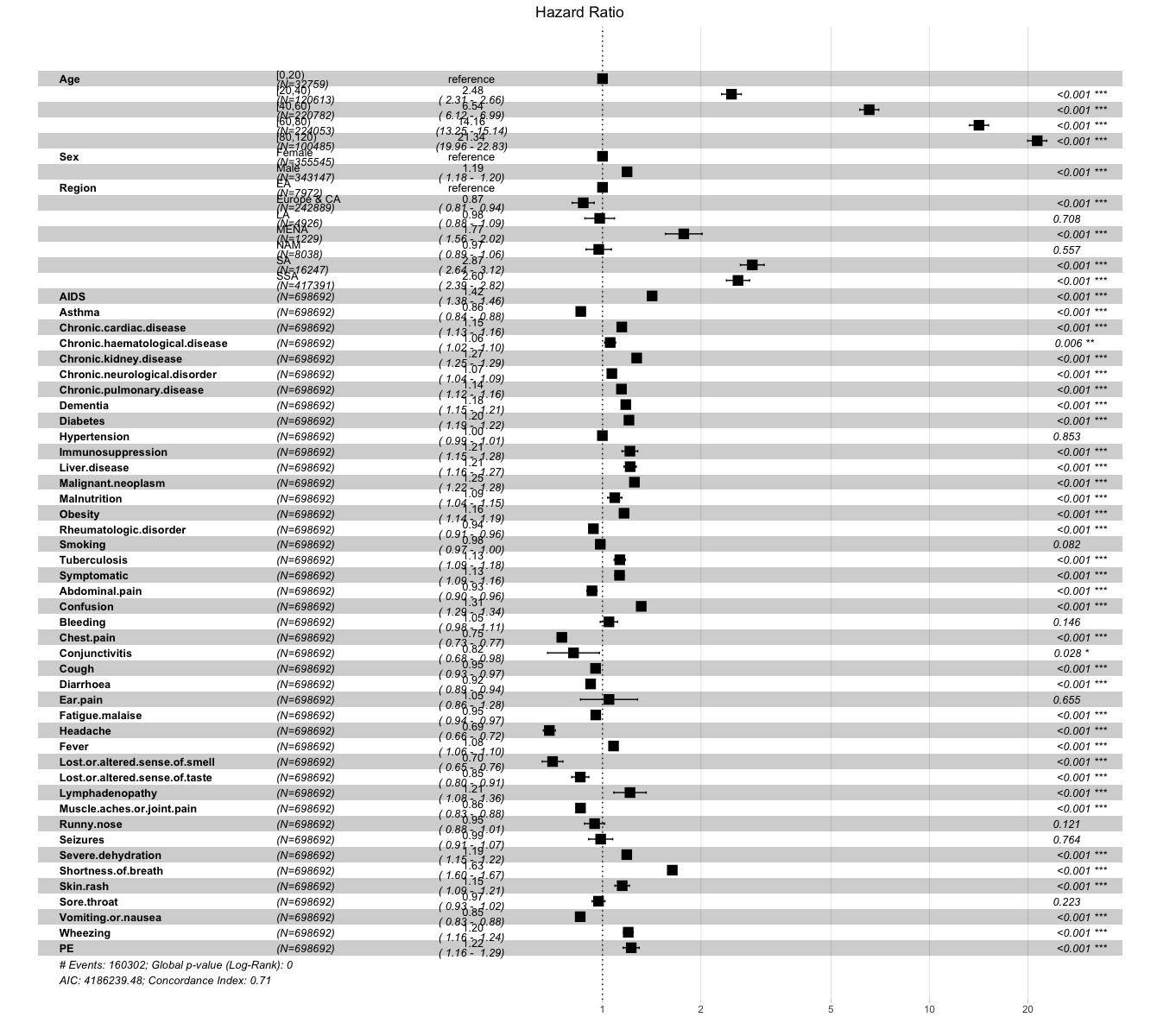}
    \caption{Adjusted Hazard Ratios for Mortality of Cox Proportional Hazards Model}
    \label{fig:CoxHR_no_lab}
\end{figure}

\noindent The Cox Proportional Hazards model without regularisation yielded a C-index of 0.71 and the forest plot shows high hazard ratios for age, certain regions of admission, and specific symptoms (Figure \ref{fig:CoxHR_no_lab} and Tables \ref{tab:HR1} and \ref{tab:HR2}). \\

\begin{table*}[t]
\caption{\textbf{Adjusted Hazard Ratios for Mortality of Cox Proportional Hazards Model}}
\small
\centering
\begin{tabular}{@{}lccc@{}}
\toprule
    \textbf{Feature} & \textbf{HR} & \textbf{95\% CI} & \textbf{P Value}\\
    \midrule
          Age &  & & \\
          \addlinespace[0.05cm]
          \hspace{0.5cm} 20-40 & 2.5 & 2.3, 2.7  & $<$0.005\\
          \addlinespace[0.05cm]
          \hspace{0.5cm} 40-60 & 6.5 & 6.1, 7.0 & $<$0.005\\
          \addlinespace[0.05cm]
          \hspace{0.5cm} 60-80 & 14.2 & 13.3, 15.1 & $<$0.005\\
          \addlinespace[0.05cm]
          \hspace{0.5cm} $>$80 & 21.3 & 20.0, 22.8 & $<$0.005\\
          \addlinespace[0.05cm]
          Sex (male) & 1.2 & 1.2, 1.2 & $<$0.005\\
          \addlinespace[0.05cm]
          \textbf{Region} &  &   &\\
          \addlinespace[0.05cm]
          \hspace{0.5cm} Sub-saharan Africa & 2.6 & 2.4, 2.8 & $<$0.005\\
          \addlinespace[0.05cm]
          \hspace{0.5cm} Europe and Central Asia & 0.9 & 0.8, 0.9 & $<$0.005\\
          \addlinespace[0.05cm]
          \hspace{0.5cm} South Asia & 2.9 & 2.6, 3.1 & $<$0.005\\
          \addlinespace[0.05cm]
          \hspace{0.5cm} North America & 1.0 & 0.9, 1.1 & 0.557\\
          \addlinespace[0.05cm]
          \hspace{0.5cm} MENA & 1.8 & 1.6, 2.0 & $<$0.005\\
          \addlinespace[0.05cm]
          \textbf{Comorbidities} &  &   &\\
          \addlinespace[0.05cm]
          \hspace{0.5cm} AIDS/HIV & 1.4 & 1.4, 1.5 & $<$0.005\\
          \addlinespace[0.05cm]
          \hspace{0.5cm} Asthma & 0.9 & 0.8, 0.9 & $<$0.005\\
          \addlinespace[0.05cm]
          \hspace{0.5cm} Chronic Cardiac Disease & 1.2 & 1.1, 1.2 & $<$0.005\\
          \addlinespace[0.05cm]
          \hspace{0.5cm} Chronic Haematological & 1.1 & 1.0, 1.1 & 0.006\\
          \addlinespace[0.05cm]
          \hspace{0.5cm} Chronic Kidney Disease & 1.3 & 1.3, 1.3 & $<$0.005\\
          \addlinespace[0.05cm]
          \hspace{0.5cm} Chronic Neurological & 1.1 & 1.0, 1.1 & $<$0.005\\
          \addlinespace[0.05cm]
          \hspace{0.5cm} Chronic Pulmonary & 1.1 & 1.1, 1.2 & $<$0.005\\
          \addlinespace[0.05cm]
          \hspace{0.5cm} Dementia & 1.2 & 1.2, 1.2 & $<$0.005\\
          \addlinespace[0.05cm]
          \hspace{0.5cm} Diabetes & 1.2 & 1.2, 1.2 & $<$0.005\\
          \addlinespace[0.05cm]
          \hspace{0.5cm} Hypertension & 1.0 & 1.0, 1.0 & 0.853\\
          \addlinespace[0.05cm]
          \hspace{0.5cm} Immunosuppression & 1.2 & 1.2, 1.3 & $<$0.005\\
          \addlinespace[0.05cm]
          \hspace{0.5cm} Liver Disease & 1.2 & 1.2, 1.3 & $<$0.005\\
          \addlinespace[0.05cm]
          \hspace{0.5cm} Malignant Neoplasm & 1.3 & 1.2, 1.3 & $<$0.005\\
          \addlinespace[0.05cm]
          \hspace{0.5cm} Malnutrition & 1.1 & 1.0, 1.2 & $<$0.005\\
          \addlinespace[0.05cm]
          \hspace{0.5cm} Obesity & 1.2 & 1.1, 1.2 & $<$0.005\\
          \addlinespace[0.05cm]
          \hspace{0.5cm} Rheumatologic & 0.9 & 0.9, 1.0 & $<$0.005\\
          \addlinespace[0.05cm]
          \hspace{0.5cm} Smoking & 1.0 & 1.0, 1.0 & 0.082\\
          \addlinespace[0.05cm]
          \hspace{0.5cm} Tuberculosis & 1.1 & 1.1, 1.2 & $<$0.005\\
          \bottomrule    
\end{tabular}
\label{tab:HR1}
\end{table*}

\begin{table*}[t]
\caption{\textbf{Adjusted Hazard Ratios for Mortality of Cox Proportional Hazards Model (continued)}}
\small
\centering
\begin{tabular}{@{}lccc@{}}
\toprule
    \textbf{Feature} & \textbf{HR} & \textbf{95\% CI} & \textbf{P Value}\\
    \midrule
          \textbf{Symptoms} & &  & \\
          \addlinespace[0.05cm]
          \hspace{0.5cm} Symptomatic & 1.1 & 1.1, 1.2 & $<$0.005\\
          \addlinespace[0.05cm]
          \hspace{0.5cm} Abdominal Pain & 0.9 & 0.9, 1.0 & $<$0.005\\
          \addlinespace[0.05cm]
          \hspace{0.5cm} Confusion & 1.3 & 1.3, 1.3 & $<$0.005\\
          \addlinespace[0.05cm]
          \hspace{0.5cm} Bleeding & 1.1 & 1.0, 1.1 & 0.146\\
          \addlinespace[0.05cm]
          \hspace{0.5cm} Chest Pain & 0.8 & 0.7, 0.8 & $<$0.005\\
          \addlinespace[0.05cm]
          \hspace{0.5cm} Conjunctivitis & 0.8 & 0.7, 1.0 & 0.028\\
          \addlinespace[0.05cm]
          \hspace{0.5cm} Cough & 1.0 & 0.9, 1.0 & $<$0.005\\
          \addlinespace[0.05cm]
          \hspace{0.5cm} Diarrhoea & 0.9 & 0.9, 0.9 & $<$0.005\\
          \addlinespace[0.05cm]
          \hspace{0.5cm} Ear Pain & 1.1 & 0.9, 1.3 & 0.655\\
          \addlinespace[0.05cm]
          \hspace{0.5cm} Fatigue & 1.0 & 0.9, 1.0 & $<$0.005\\
          \addlinespace[0.05cm]
          \hspace{0.5cm} Headache & 0.7 & 0.7, 0.7 & $<$0.005\\
          \addlinespace[0.05cm]
          \hspace{0.5cm} Fever & 1.1 & 1.1, 1.1 & $<$0.005\\
          \addlinespace[0.05cm]
          \hspace{0.5cm} Lost Sense of Smell & 0.7 & 0.7, 0.8 & $<$0.005\\
          \addlinespace[0.05cm]
          \hspace{0.5cm} Lost Sense of Taste & 0.9 & 0.8, 0.9 & $<$0.005\\
          \addlinespace[0.05cm]
          \hspace{0.5cm} Lymphadenopathy & 1.2 & 1.1, 1.4 & $<$0.005\\
          \addlinespace[0.05cm]
          \hspace{0.5cm} Muscle/Joint Pain & 0.9 & 0.8, 0.9 & $<$0.005\\
          \addlinespace[0.05cm]
          \hspace{0.5cm} Runny Nose & 1.0 & 0.9, 1.0 & 0.121\\
          \addlinespace[0.05cm]
          \hspace{0.5cm} Seizures & 1.0 & 0.9, 1.1 & 0.764\\
          \addlinespace[0.05cm]
          \hspace{0.5cm} Severe Dehydration & 1.2 & 1.2, 1.2 & $<$0.005\\
          \addlinespace[0.05cm]
          \hspace{0.5cm} Shortness of Breath & 1.6 & 1.6, 1.7 & $<$0.005\\
          \addlinespace[0.05cm]
          \hspace{0.5cm} Skin Rash & 1.2 & 1.1, 1.2 & $<$0.005\\
          \addlinespace[0.05cm]
          \hspace{0.5cm} Sore Throat & 1.0 & 0.9, 1.0 & 0.223\\
          \addlinespace[0.05cm]
          \hspace{0.5cm} Vomiting & 0.9 & 0.8, 0.9 & $<$0.005\\
          \addlinespace[0.05cm]
          \hspace{0.5cm} Wheezing & 1.2 & 1.2, 1.2 & $<$0.005\\
          \addlinespace[0.05cm]
          PE & 1.2 & 1.2, 1.3 & $<$0.005\\
            \bottomrule    
\end{tabular}
\label{tab:HR2}
\end{table*}

\begin{figure*}[t]
    \centering
    \begin{subfigure}[t]{0.65\textwidth}
        \centering
        \includegraphics[width=\linewidth]{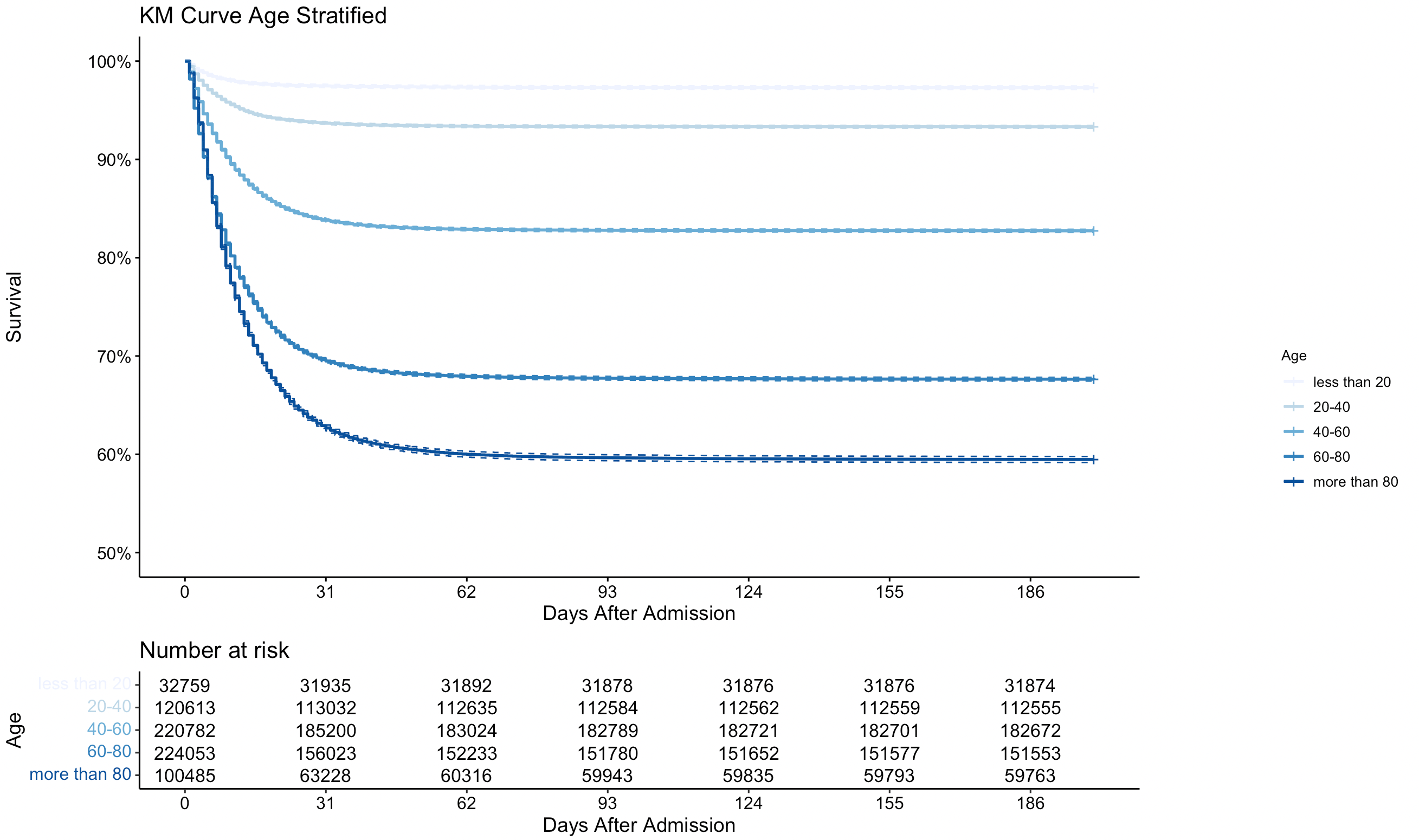} 
        \caption{By Age} \label{fig:KM_Age}
    \end{subfigure}
    \hfill
    \begin{subfigure}[t]{0.65\textwidth}
        \centering
        \includegraphics[width=\linewidth]{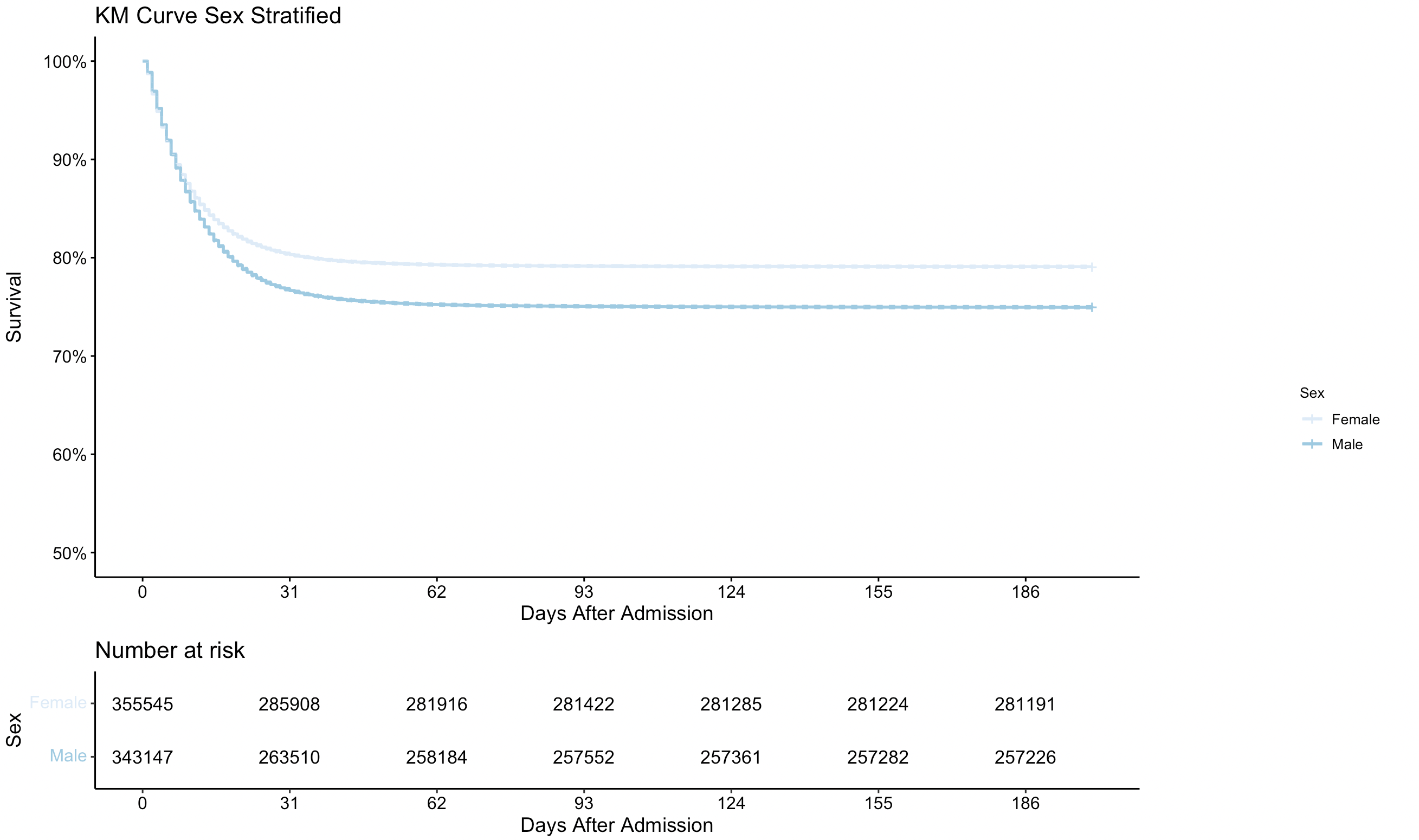} 
        \caption{By Sex} \label{fig:KM_Sex}
    \end{subfigure}
    \hfill
    \begin{subfigure}[t]{0.65\textwidth}
        \centering
        \includegraphics[width=\linewidth]{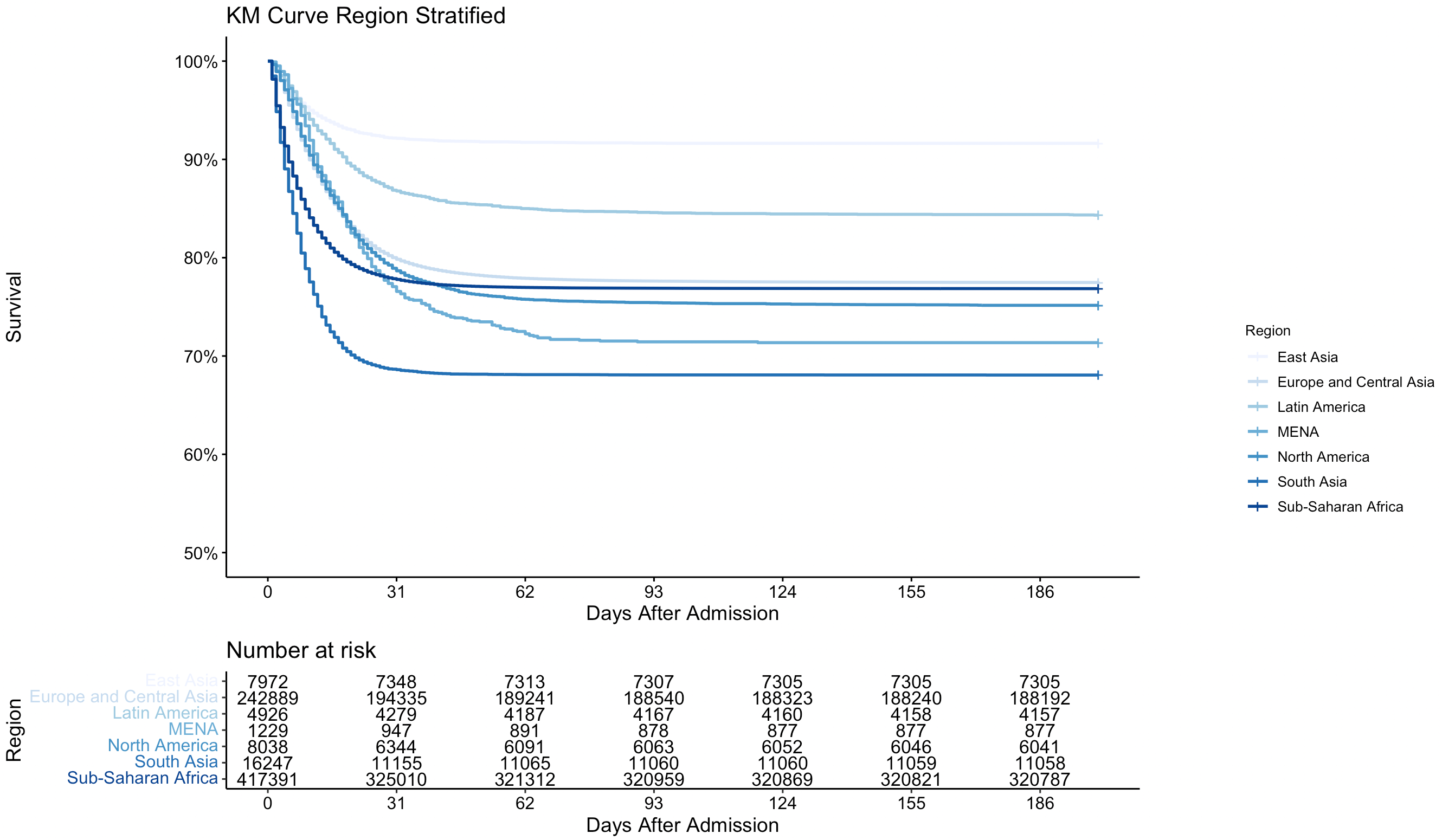} 
        \caption{By Region of Admission} \label{fig:KM_Region}
    \end{subfigure}
    \caption{Kaplan-Meier Survival Curve For COVID-19 Patients Stratified}
    \label{fig:KM}
\end{figure*}

\noindent The Kaplan-Meier curves for risk stratification across age, sex, and region groups show clear difference in risk with older men and those in South Asia and the Middle East with the lowest rates of survival (Figure \ref{fig:KM}). \\

\noindent Tables \ref{tab:PE_UKSpain}, \ref{tab:PE_UK}, and \ref{tab:PE_Spain} show superior performance of the XGBoost model across all 3 test sets. Similarly, XGBoost maintains sensitive and accurate prediction of death compared to other alternative models (Table \ref{tab:Death_All}). The validation scores are for the combined UK and Spain set. \\

\begin{table*}[t]
\caption{\textbf{Prediction Model Results for PE on Test Set with UK and Spain (F1-w is the weighted F1 score)}}
\small
\centering
\begin{tabular}{@{}lcccccc@{}}
\toprule
    \textbf{Models} & \textbf{Validation AUC} & \textbf{AUC} & \textbf{Accuracy} & \textbf{F1-w} & \textbf{Sensitivity}\\
    \midrule
\textbf{No undersampling} &  &  & &&\\
\textbf{No threshold} &  &  & &&\\
\midrule
Logistic Regression & 72.5 & 71.0 & 64.0 & 76.5 & 69.3 \\
\midrule
LDA & 72.2 & 70.6 & 98.3 & 97.4 & 0.0 \\
\midrule
Naive Bayes & 70.4 & 69.1 & 98.1 & 97.4 & 0.9 \\
\midrule
Random Forest & 73.6 & 73.5 & 65.4 & 77.5 & 69.7 \\
\midrule
Stacking Ensemble & 63.0 & 67.3 & 65.5 & 77.6 & 69.1 \\
\midrule
Ensemble & 73.0 & 71.8 & 63.7 & 76.2 & 70.8 \\
\midrule
Ensemble (XGBoost) & 73.6 & 73.8 & 64.5 & 76.9 & 70.0 \\
\midrule
\textbf{XGBoost} & \textbf{75.6} & \textbf{75.9} & \textbf{72.3} & \textbf{82.3} & \textbf{67.5} \\
    \midrule
\textbf{No undersampling} &  &   & &&\\
\textbf{With threshold} &  &   & &&\\
\midrule
Logistic Regression & 72.5 & 66.7 & 63.0 & 75.7 & 70.5 \\
\midrule
LDA & 72.2 & 66.5 & 66.7 & 78.5 & 66.3 \\
\midrule
Naive Bayes & 70.4 & 65.3 & 61.8 & 74.9 & 68.9 \\
\midrule
Random Forest & 73.6 & 66.6 & 71.4 & 81.8 & 61.5 \\
\midrule
XGBoost & 73.8 & 67.3 & 67.9 & 79.3 & 66.7 \\
\midrule
\textbf{With undersampling} &   &  & &&\\
\textbf{No threshold} &  &   & &&\\
\midrule
Logistic Regression & 72.4 & 71.0 & 63.9 & 76.5 & 69.1 \\
\midrule
LDA & 72.2 & 70.6 & 95.9 & 96.4 & 9.8 \\
\midrule
Naive Bayes & 70.4 & 69.0 & 82.7 & 89.0 & 34.2 \\
\midrule
Random Forest & 74.3 & 73.9 & 68.3 & 79.6 & 66.6 \\
\midrule
Stacking Ensemble & 64.5 & 67.6 & 66.8 & 78.6 & 68.4 \\
\midrule
XGBoost & 73.8 & 73.7 & 66.1 & 78.1 & 69.0 \\
\midrule
\textbf{With undersampling} &   &  & &&\\
\textbf{With threshold} &  &   & &&\\
\midrule
Logistic Regression & 72.4 & 66.3 & 64.7 & 77.1 & 67.9 \\
\midrule
LDA & 72.2 & 66.7 & 63.2 & 75.9 & 70.3 \\
\midrule
Naive Bayes & 70.4 & 65.2 & 61.5 & 74.6 & 68.9 \\
\midrule
Random Forest & 74.3 & 73.9 & 68.3& 79.6 & 66.6 \\
\midrule
XGBoost & 73.8 & 67.3 & 62.4 & 75.3 & 72.5 \\
  \bottomrule                          
\end{tabular}
\label{tab:PE_UKSpain}
\end{table*}

\begin{table*}[t]
\caption{\textbf{Prediction Model Results for PE on UK Test Set (F1-w is the weighted F1 score)}}
\small
\centering
\begin{tabular}{@{}lccccc@{}}
\toprule
    \textbf{Models} & \textbf{Validation AUC} & \textbf{AUC} & \textbf{Accuracy} & \textbf{F1-w} & \textbf{Sensitivity}\\
    \midrule
\textbf{No undersampling} &  &  & &&\\
\textbf{No threshold} &  &  & &&\\
\midrule
Logistic Regression & 72.5 & 69.4 & 64.9 & 77.3 & 65.5 \\
\midrule
LDA & 72.2 & 69.2 & 98.4 & 97.6 & 0.0 \\
\midrule
Naive Bayes & 70.4 & 67.2 & 98.3 & 97.6 & 0.5 \\
\midrule
Random Forest & 73.6 & 71.2 & 65.6 & 77.8 & 66.0 \\
\midrule
Stacking Ensemble & 63.0 & 65.7 & 66.1 & 78.2 & 65.2 \\
\midrule
Ensemble & 73.0 & 70.3 & 64.7 & 77.1 & 67.0 \\
\midrule
Ensemble (XGBoost) & 73.6 & 71.6 & 64.9 & 77.3 & 66.1 \\
\midrule
\textbf{XGBoost} & \textbf{75.6} & \textbf{74.5} & \textbf{73.4} & \textbf{83.2} & \textbf{63.5} \\
    \midrule
\textbf{No undersampling} &  &   & &&\\
\textbf{With threshold} &  &  & &&\\
\midrule
Logistic Regression & 72.5 & 65.3 & 63.9 & 76.5 & 66.8 \\
\midrule
LDA & 72.2 & 65.3 & 68.7 & 80.0 & 61.9 \\
\midrule
Naive Bayes & 70.4 & 63.6 & 61.9 & 75.0 & 65.5 \\
\midrule
Random Forest & 73.6 & 64.5 & 71.9 & 82.2 & 56.9 \\
\midrule
XGBoost & 73.8 & 65.5 & 68.5 & 79.9 & 62.5 \\
\midrule
\textbf{With undersampling} &  &  & &&\\
\textbf{No threshold} &  &  & &&\\
\midrule
Logistic Regression & 72.4 & 69.3 & 64.8 & 77.2 & 65.2 \\
\midrule
LDA & 72.2 & 69.2 & 97.0 & 97.0 & 4.8 \\
\midrule
Naive Bayes & 70.4 & 67.2 & 83.3 & 89.5 & 29.7 \\
\midrule
Random Forest & 74.3 & 71.7 & 68.8 & 80.1 & 62.4 \\
\midrule
Stacking Ensemble & 64.5 & 65.9 & 67.5 & 79.2 & 64.3 \\
\midrule
XGBoost & 73.8 & 71.7 & 66.8 & 78.6 & 65.1 \\
\midrule
\textbf{With undersampling} &   &  & &&\\
\textbf{With threshold} &  &   & &&\\
\midrule
Logistic Regression & 72.4 & 64.8 & 65.6 & 77.8 & 64.0 \\
\midrule
LDA & 72.2 & 65.5 & 64.5 & 77.0 & 66.6 \\
\midrule
Naive Bayes & 70.4 & 63.5 & 61.6 & 74.8 & 65.5 \\
\midrule
Random Forest & 74.3 & 65.5 & 69.3 & 80.5 & 61.6 \\
\midrule
XGBoost & 73.8 & 65.9 & 62.8 & 75.7 & 69.1 \\
  \bottomrule                          
\end{tabular}
\label{tab:PE_UK}
\end{table*}

\begin{table*}[t]
\caption{\textbf{Prediction Model Results for PE on Spain Test Set (F1-w is the weighted F1 score)}}
\small
\centering
\begin{tabular}{@{}lccccc@{}}
\toprule
    \textbf{Models} & \textbf{Validation AUC} & \textbf{AUC} & \textbf{Accuracy} & \textbf{F1-w} & \textbf{Sensitivity}\\
    \midrule
\textbf{No undersampling} &  &    & &&\\
\textbf{No threshold} &  &  &   &&\\
\midrule
Logistic Regression & 72.5 & 74.5 & 48.5 & 61.6 & 96.5 \\
\midrule
LDA & 72.2 & 73.6 & 96.2 & 94.4 & 0.0 \\
\midrule
Naive Bayes & 70.4 & 77.6 & 95.7 & 94.3 & 3.5 \\
\midrule
Random Forest & 73.6 & 78.7 & 62.1 & 73.3 & 95.7 \\
\midrule
Stacking Ensemble & 63.0 & 75.7 & 56.4 & 68.6 & 96.5 \\
\midrule
Ensemble & 73.0 & 74.4 & 46.5 & 60.0 & 95.4 \\
\midrule
Ensemble (XGBoost) & 73.6 & 79.8 & 57.8 & 69.8 & 95.4 \\
\midrule
\textbf{XGBoost} & \textbf{75.6} & \textbf{78.9} & \textbf{9.6} & \textbf{67.2} & \textbf{95.7} \\
    \midrule
\textbf{No undersampling} &  &    & &&\\
\textbf{With threshold} &  &    & &&\\
\midrule
Logistic Regression & 72.5 & 71.3 & 48.1 & 61.2 & 96.5 \\
\midrule
LDA & 72.2 & 64.2 & 33.5 & 45.8 & 97.4 \\
\midrule
Naive Bayes & 70.4 & 76.3 & 60.9 & 72.4 & 93.0 \\
\midrule
Random Forest & 73.6 & 78.4 & 64.1 & 74.9 & 93.9 \\
\midrule
XGBoost & 73.8 & 76.0 & 57.1 & 69.2 & 96.5 \\
\midrule
\textbf{With undersampling} &    &  & &&\\
\textbf{No threshold} &  &    & &&\\
\midrule
Logistic Regression & 72.4 & 74.6 & 49.5 & 62.6 & 96.5 \\
\midrule
LDA & 72.2 & 73.9 & 78.7 & 85.0 & 44.3 \\
\midrule
Naive Bayes & 70.4 & 77.6 & 71.8 & 80.5 & 66.1 \\
\midrule
Random Forest &  74.3 &  80.2 &  58.7 & 70.6 & 96.5 \\
\midrule
Stacking Ensemble & 64.5 & 75.0 & 55.1 & 67.5 & 96.5 \\
\midrule
XGBoost & 73.8 & 79.3 & 54.9 & 67.4 & 96.5 \\
\midrule
\textbf{With undersampling} &   &  & &&\\
\textbf{With threshold} &  &   & &&\\
\midrule
Logistic Regression & 72.4 & 71.8 & 49.7 & 62.7 & 95.7 \\
\midrule
LDA & 72.2 & 68.2 & 42.0 & 55.2 & 96.5 \\
\midrule
Naive Bayes & 70.4 & 76.4 & 61.1 & 72.5 & 93.0 \\
\midrule
Random Forest & 74.3 & 76.9 & 58.7 & 70.6 & 96.5 \\
\midrule
XGBoost & 73.8 & 74.7 & 54.4 & 66.9 & 96.5 \\
  \bottomrule                          
\end{tabular}
\label{tab:PE_Spain}
\end{table*}

\begin{table*}[t]
\caption{\textbf{Death Prediction Model Results for Test Set (F1-w is the weighted F1 score)}}
\small
\centering
\begin{tabular}{@{}lccccc@{}}
\toprule
    \textbf{Models} & \textbf{Validation AUC} & \textbf{AUC} & \textbf{Accuracy} & \textbf{F1-w} & \textbf{Sensitivity}\\
    \midrule
\textbf{No undersampling} &  &  & &&\\
\textbf{No threshold} &  &  & &&\\
\midrule
Logistic Regression & 73.2 & 72.9 & 66.2 & 68.9 & 68.4 \\
\midrule
LDA & 73.1 & 72.9 & 78.0 & 71.1 & 7.8 \\
\midrule
Naive Bayes & 71.3 & 71.1 & 74.9 & 72.5 & 23.7 \\
\midrule
Random Forest & 74.1 & 73.9 & 65.5 & 68.4 & 71.5 \\
\midrule
Stacking Ensemble & 74.1 & 73.9 & 65.5 & 68.4 & 71.5 \\
\midrule
Ensemble & 73.3 & 73.1 & 65.8 & 68.6 & 69.4 \\
\midrule
Ensemble (XGBoost) & 74.4 & 74.3 & 65.1 & 68.1 & 73.0 \\
\midrule
\textbf{XGBoost} & \textbf{74.4} & \textbf{74.2} & \textbf{65.3} & \textbf{68.2} & \textbf{72.7} \\
    \midrule
\textbf{No undersampling} &  &    & &&\\
\textbf{With threshold} &  &    & &&\\
\midrule
Logistic Regression & 73.2 & 67.0 & 63.7 & 66.8 & 72.9 \\
\midrule
LDA & 73.1 & 67.1 & 63.8 & 66.9 & 72.8 \\
\midrule
Naive Bayes & 71.3 & 66.9 & 62.6 & 65.8 & 74.7 \\
\midrule
Random Forest & 74.1 & 67.6 & 65.9 & 68.8 & 70.6 \\
\midrule
Ensemble & 73.3 & 67.1 & 62.8 & 66.0 & 74.8 \\
\midrule
Ensemble (XGBoost) & 74.4 & 67.9 & 66.3 & 69.1 & 70.7 \\
\midrule
XGBoost & 74.4 & 67.9 & 66.1 & 69.0 & 71.1 \\
  \bottomrule                          
\end{tabular}
\label{tab:Death_All}
\end{table*}

\noindent The model also maintains high predictive performance across various subgroups of the patient population stratified across sex and age (Tables \ref{tab:Stratified_Results_PE} and \ref{tab:Stratified_Results_Death}). \\

\noindent To further evaluate our model, we test it on held-out test data with specific patient population subgroups like men, women, and different age groups as can be seen in Tables \ref{tab:Stratified_Results_PE} and \ref{tab:Stratified_Results_Death}. Our model shows reliable prediction for PE and mortality in both men and women without a significant difference in performance for each group, whereas for age groups there is greater variation in results as compared to sex differences but it remains relatively consistent in predictive performance. \\

\begin{table*}[t]
\caption{\textbf{Prediction Model Results Stratified Across Sex and Age Groups for PE (F1-w is the weighted F1 score)}}
\small
\centering
\begin{tabular}{@{}lcccc@{}}
\toprule
    \textbf{Models} & \textbf{AUC} & \textbf{Accuracy} & \textbf{F1-w} & \textbf{Sensitivity}\\
    \midrule
\textbf{Sex} &  & &&\\
\midrule
\textbf{Male}   &  & &&\\
\midrule
Logistic Regression   & 71.1 & 55.9 & 69.9 & 77.7 \\
\midrule
XGBoost   & 76.0 & 68.0 & 79.1 & 73.0 \\
    \midrule
\textbf{Female}   &  & &&\\
\midrule
Logistic Regression   & 69.3 & 73.7 & 83.7 & 53.4 \\
\midrule
XGBoost   & 74.3 & 77.3 & 86.0 & 57.2 \\
\midrule
\textbf{Age}   &  & &&\\
\midrule
\textbf{20-40}   &  & &&\\
\midrule
Logistic Regression   & 79.0 & 65.8 & 78.0 & 81.0 \\
\midrule
XGBoost   & 78.2 & 74.1 & 83.5 & 68.8 \\
\midrule
\textbf{40-60}   &  & &&\\
\midrule
Logistic Regression   & 65.7 & 48.7 & 63.2 & 75.9 \\
\midrule
XGBoost  & 74.0 & 59.0 & 71.9 & 78.2 \\
\midrule
\textbf{60-80}   &  & &&\\
\midrule
Logistic Regression   & 69.6 & 58.6 & 72.0 & 71.5 \\
\midrule
XGBoost   & 72.3 & 66.1 & 77.6 & 67.7 \\
  \bottomrule                          
\end{tabular}
\label{tab:Stratified_Results_PE}
\end{table*}

\begin{table*}[t]
\caption{\textbf{Prediction Model Results Stratified Across Sex and Age Groups for Death (F1-w is the weighted F1 score)}}
\small
\centering
\begin{tabular}{@{}lcccc@{}}
\toprule
    \textbf{Models} & \textbf{AUC} & \textbf{Accuracy} & \textbf{F1-w} & \textbf{Sensitivity}\\
    \midrule
\textbf{Sex} &  & &&\\
\midrule
\textbf{Male}  &  & &&\\
\midrule
Logistic Regression   & 72.4 & 64.4 & 67.0 & 71.2 \\
\midrule
XGBoost   & 74.0 & 64.2 & 66.8 & 74.6 \\
    \midrule
\textbf{Female}   & & &&\\
\midrule
Logistic Regression   & 72.9 & 67.9 & 70.8 & 65.1 \\
\midrule
XGBoost   & 74.0 & 66.3 & 69.6 & 70.4 \\
\midrule
\textbf{Age}   &  & &&\\
\midrule
\textbf{20-40}   &  & &&\\
\midrule
Logistic Regression   & 67.8 & 93.4 & 90.2 & 0.0 \\
\midrule
XGBoost   & 70.2 & 93.1 & 90.5 & 4.2 \\
\midrule
\textbf{40-60}   &  & &&\\
\midrule
Logistic Regression   & 62.9 & 77.8 & 76.2 & 18.1 \\
\midrule
XGBoost  & 65.0 & 74.3 & 75.2 & 32.1 \\
\midrule
\textbf{60-80}   &  & &&\\
\midrule
Logistic Regression   & 61.7 & 47.6 & 46.4 & 85.8 \\
\midrule
XGBoost   & 63.6 & 46.9 & 44.9 & 89.2 \\
  \bottomrule                             
\end{tabular}
\label{tab:Stratified_Results_Death}
\end{table*}

\begin{figure}[H]
    \centering
    \includegraphics[width=1.0\linewidth]{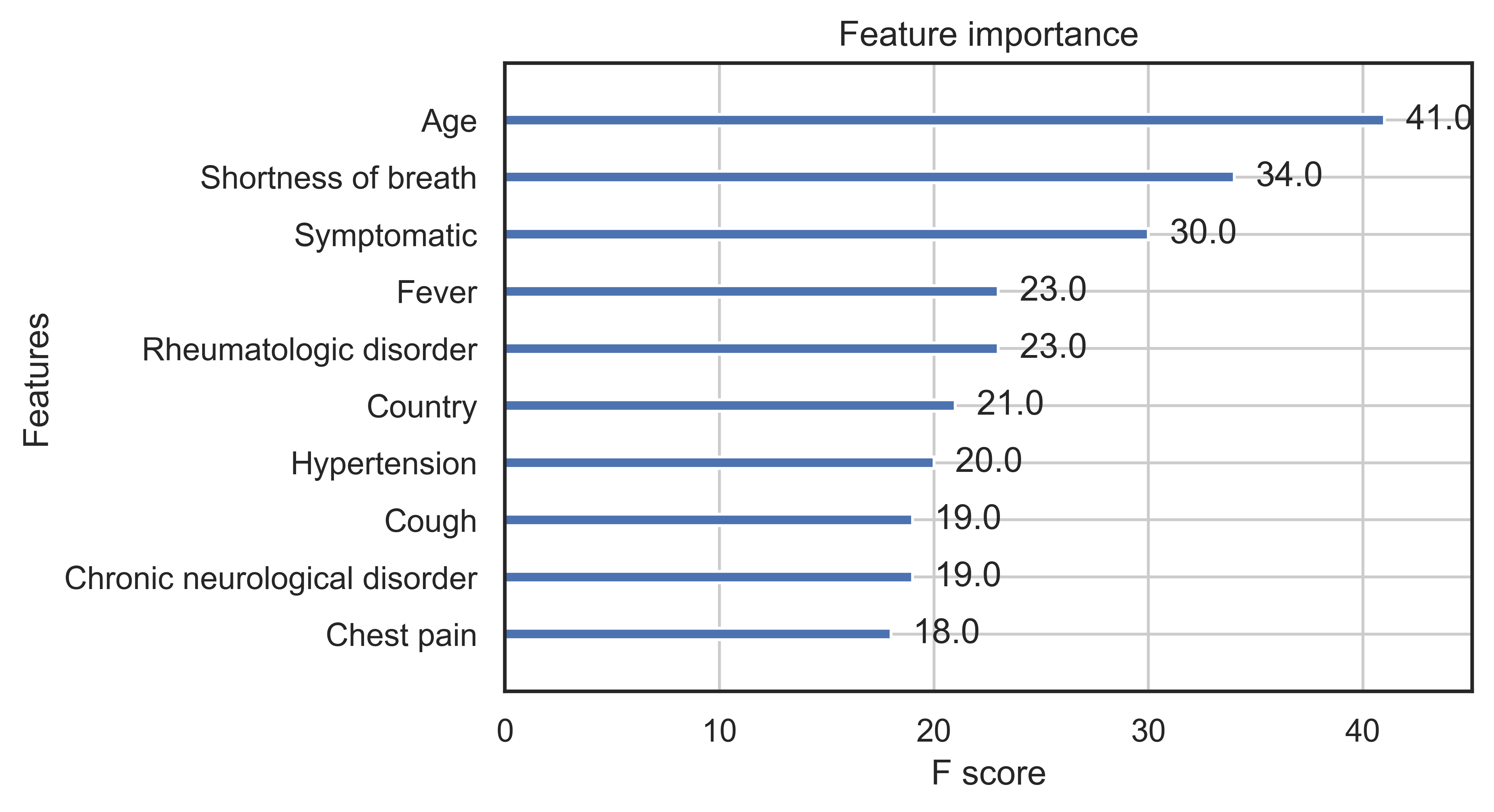}
    \caption{Feature Importance From XGBoost PE Prediction Model Using F1-Score Gain Method (Average Contribution of Each Feature to Predictive Performance)}
    \label{fig:PE_FI}
\end{figure}

\begin{figure}[h]
    \centering
    \includegraphics[width=1.0\linewidth]{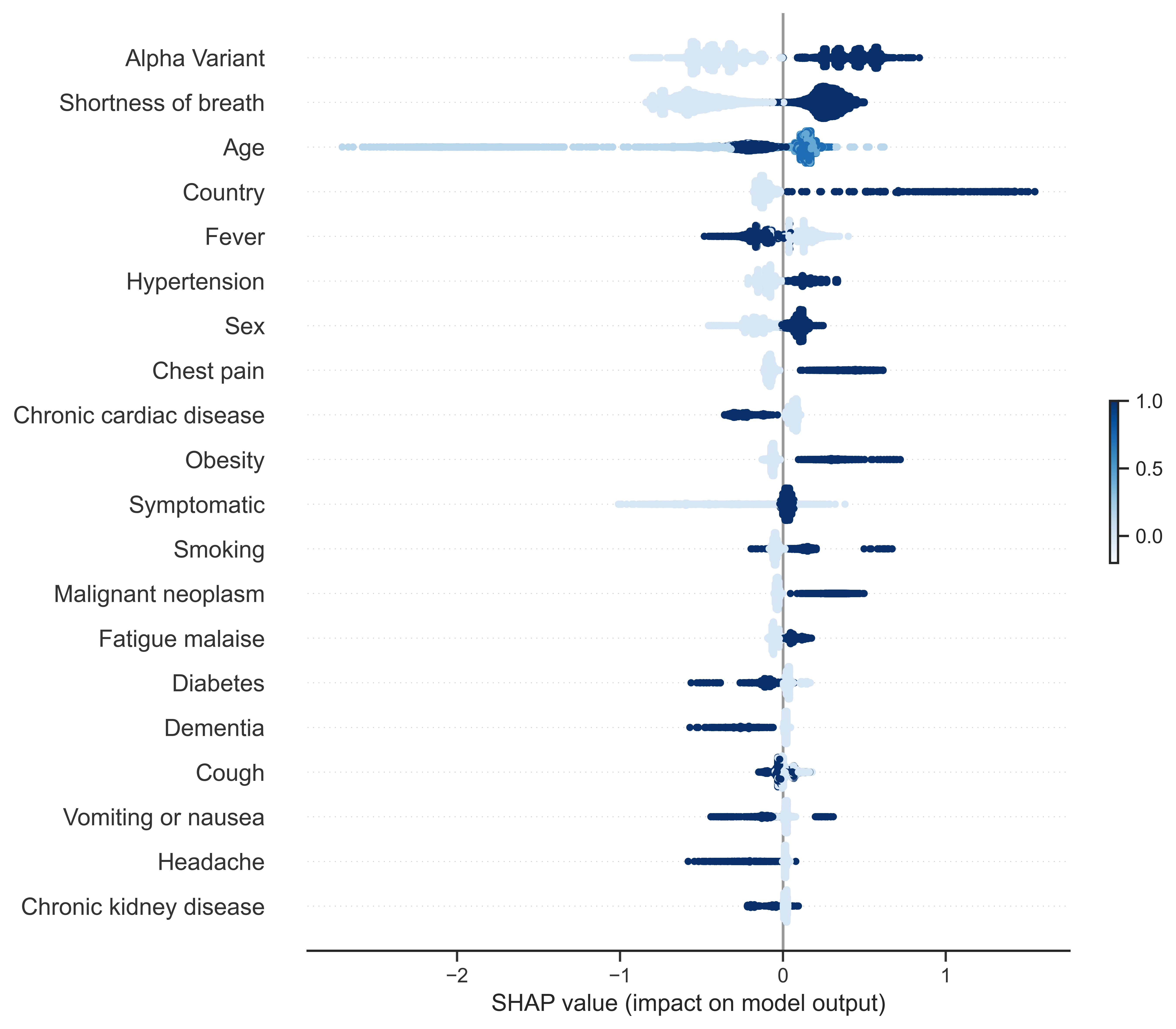}
    \caption{XGBoost Feature Importance with SHAP for PE. The values in the legend being higher or darker colour in the plot correspond to higher values of that feature contributing to the prediction either for stronger positive prediction (more colour points for the feature on the right side of the vertical line) or stronger negative prediction of outcome otherwise.}
    \label{fig:PE_Shap}
\end{figure}

\begin{figure}[H]
    \centering
    \includegraphics[width=1.0\linewidth]{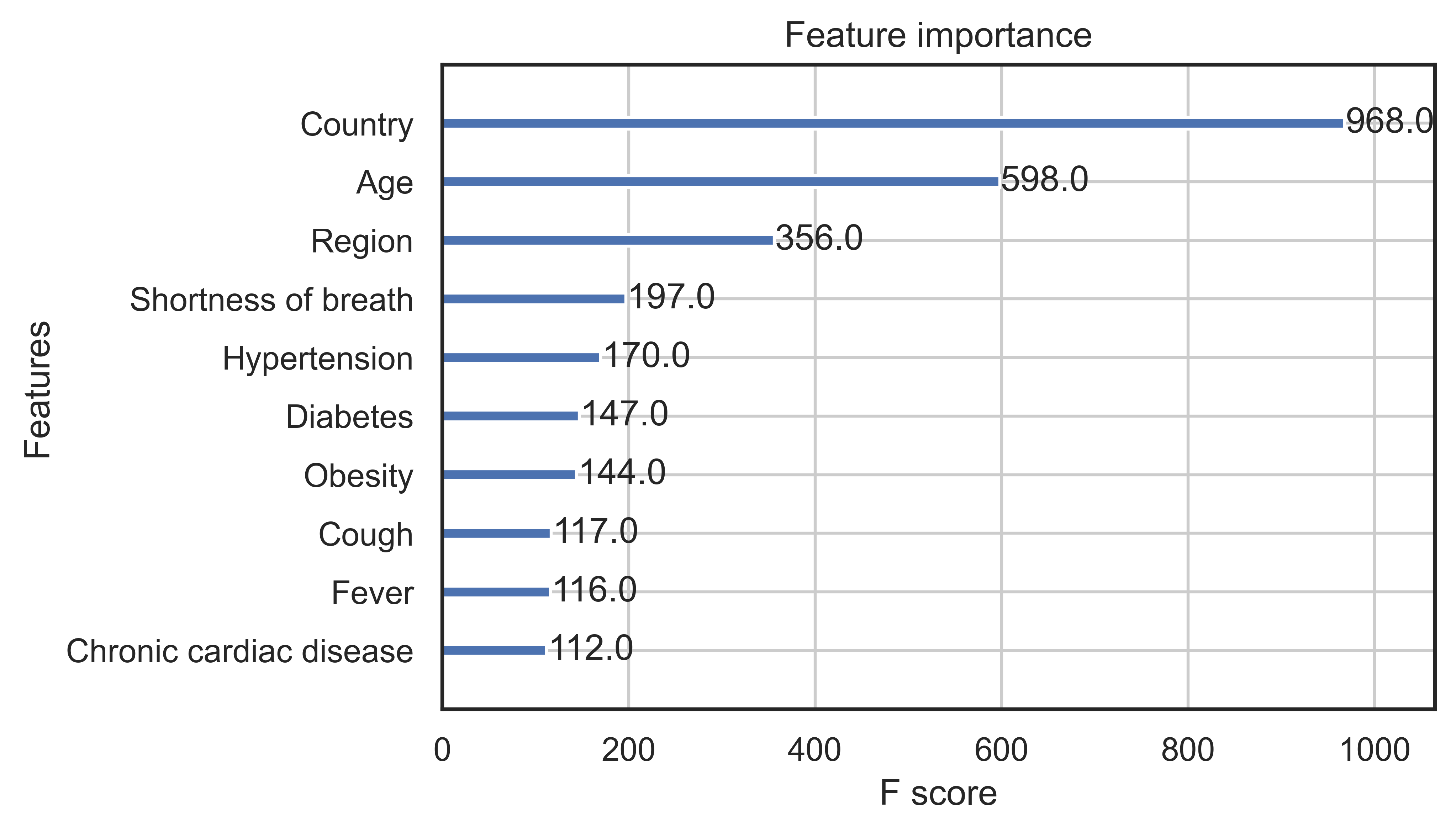}
    \caption{Feature Importance From XGBoost Mortality Prediction Model Using F1-Score Gain Method (Average Contribution of Each Feature to Predictive Performance)}
    \label{fig:Death_FI}
\end{figure}

\begin{figure}[h]
    \centering
    \includegraphics[width=1.0\linewidth]{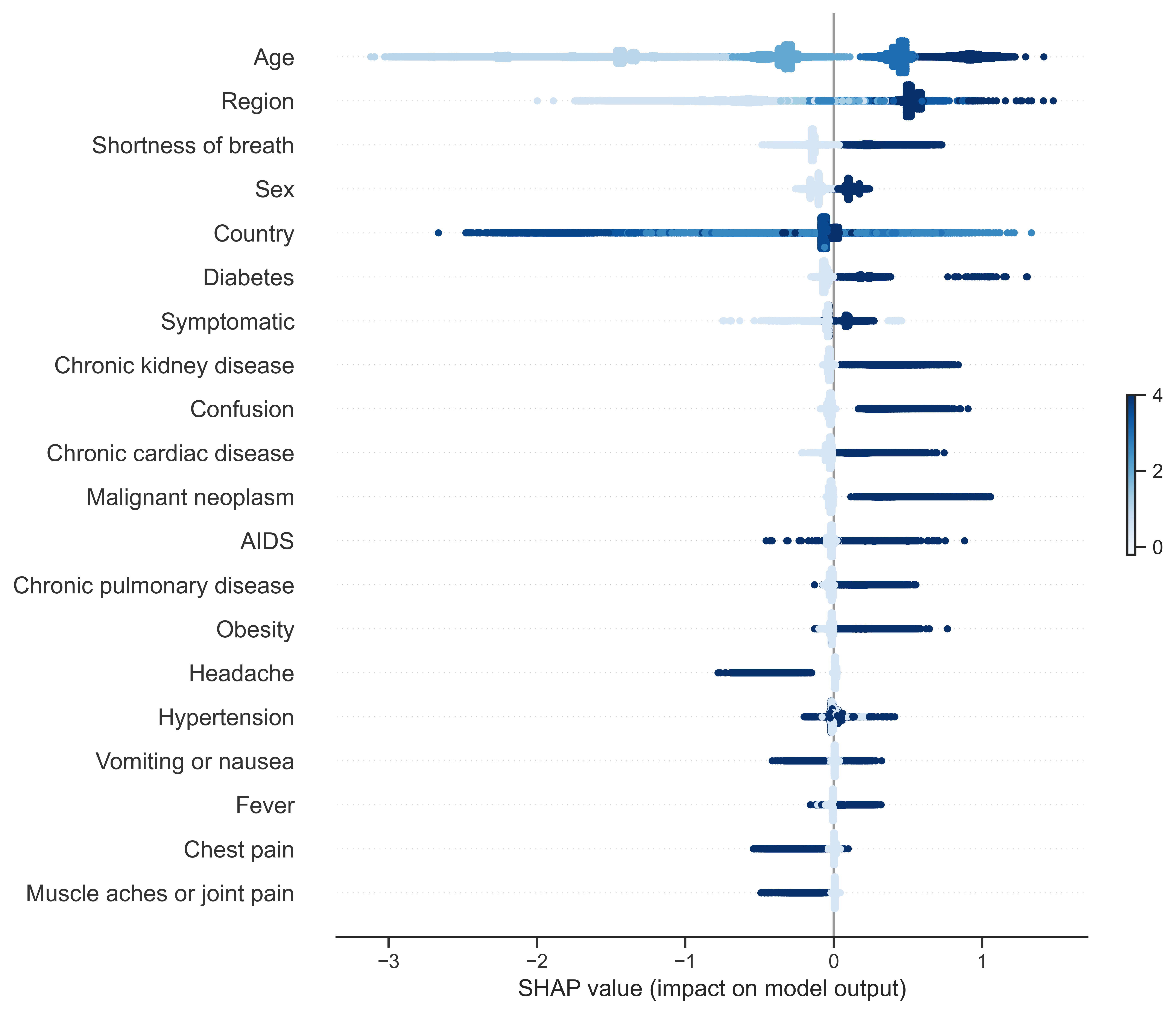}
    \caption{XGBoost Feature Importance with SHAP for Mortality. The values in the legend being higher or darker colour in the plot correspond to higher values of that feature contributing to the prediction either for stronger positive prediction (more colour points for the feature on the right side of the vertical line) or stronger negative prediction of outcome otherwise.}
    \label{fig:Death_Shap}
\end{figure}

\begin{figure}[h]
    \centering
    \includegraphics[width=1.0\linewidth]{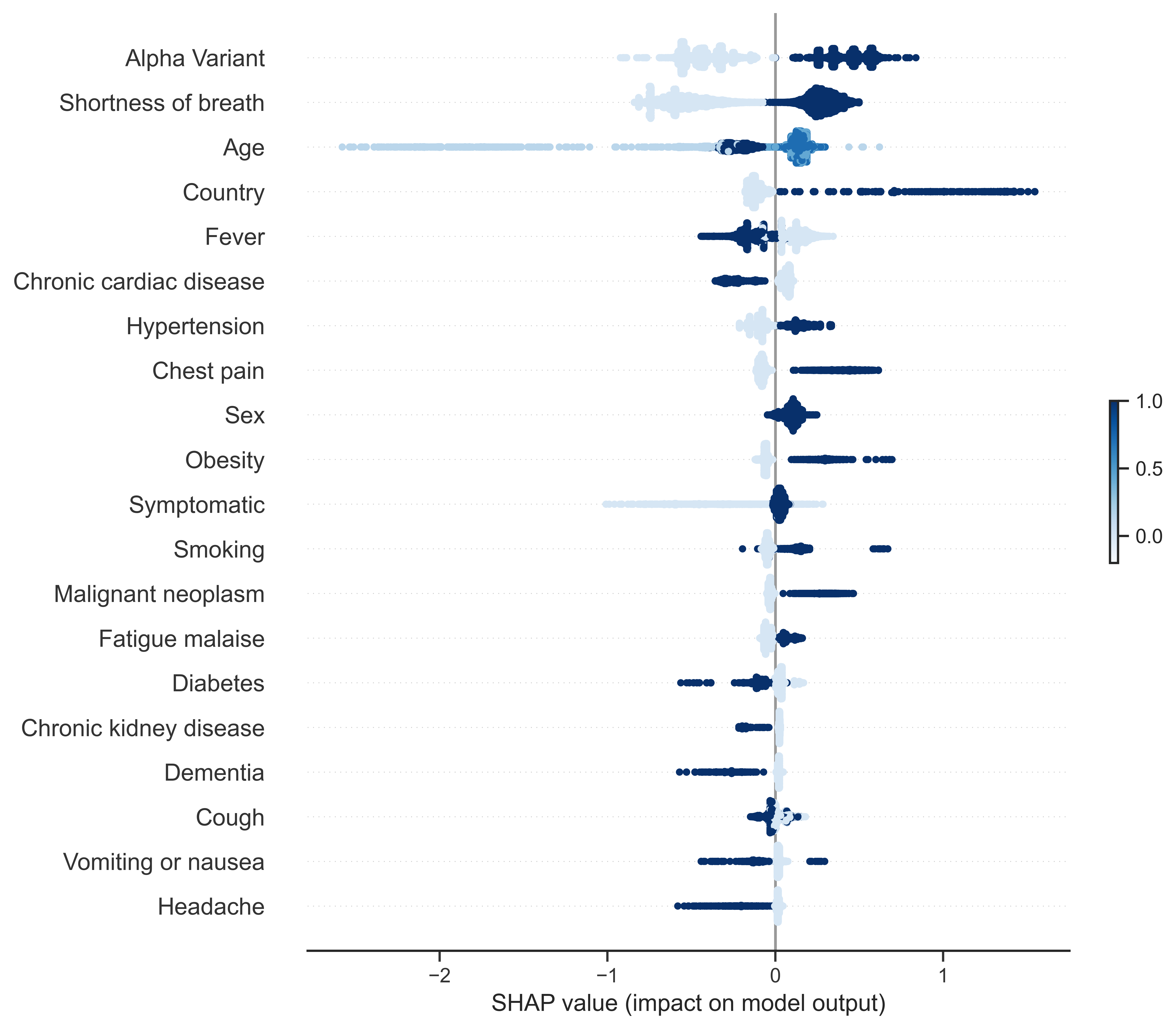}
    \caption{XGBoost Feature Importance with SHAP for PE (only Men). The values in the legend being higher or darker colour in the plot correspond to higher values of that feature contributing to the prediction either for stronger positive prediction (more colour points for the feature on the right side of the vertical line) or stronger negative prediction of outcome otherwise.}
    \label{fig:PE_Shap_Men}
\end{figure}

\begin{figure}[h]
    \centering
    \includegraphics[width=1.0\linewidth]{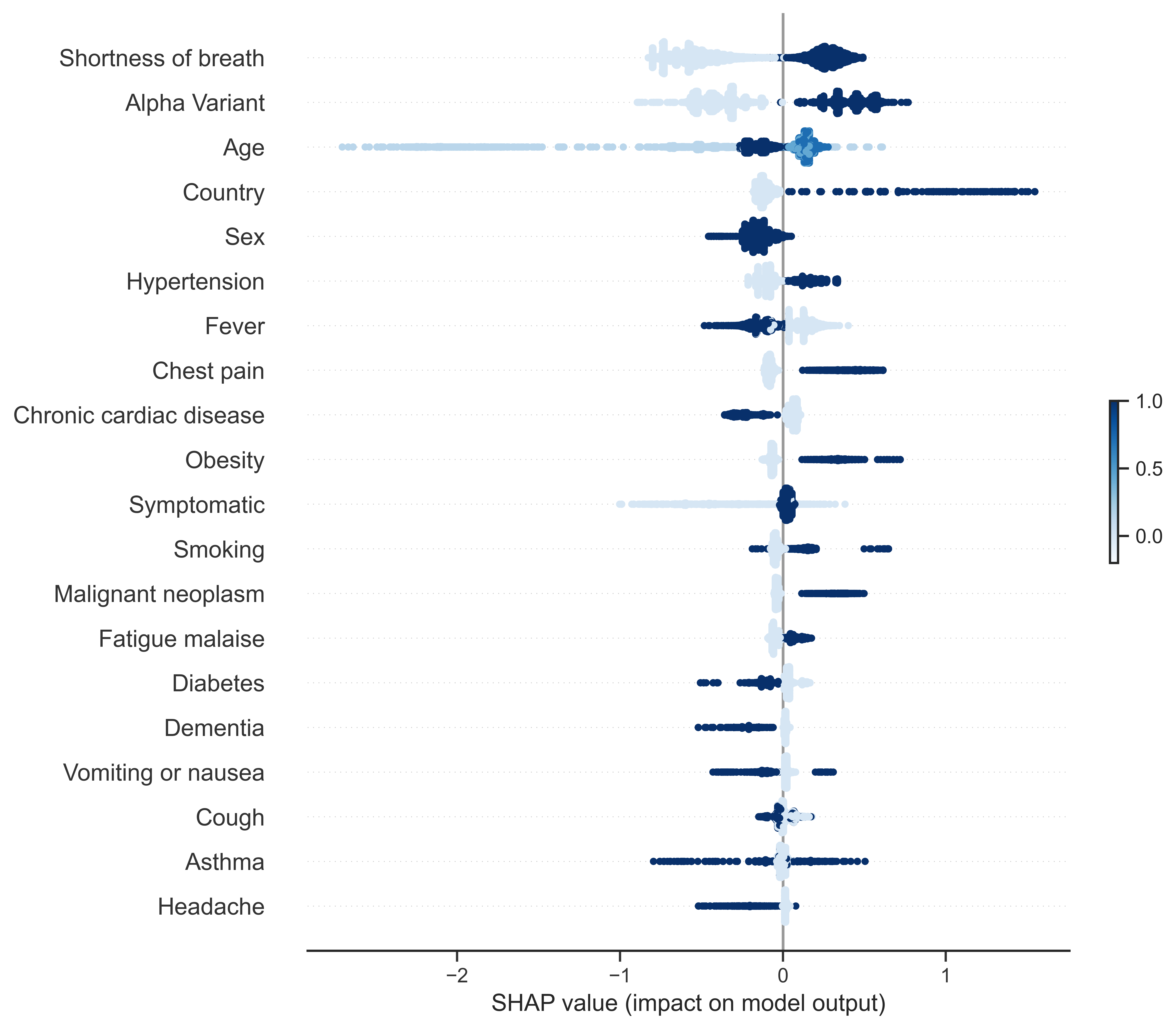}    \caption{XGBoost Feature Importance with SHAP for PE (only Women). The values in the legend being higher or darker colour in the plot correspond to higher values of that feature contributing to the prediction either for stronger positive prediction (more colour points for the feature on the right side of the vertical line) or stronger negative prediction of outcome otherwise.}
    \label{fig:PE_Shap_Women}
\end{figure}

\begin{figure}[h]
    \centering
    \includegraphics[width=1.0\linewidth]{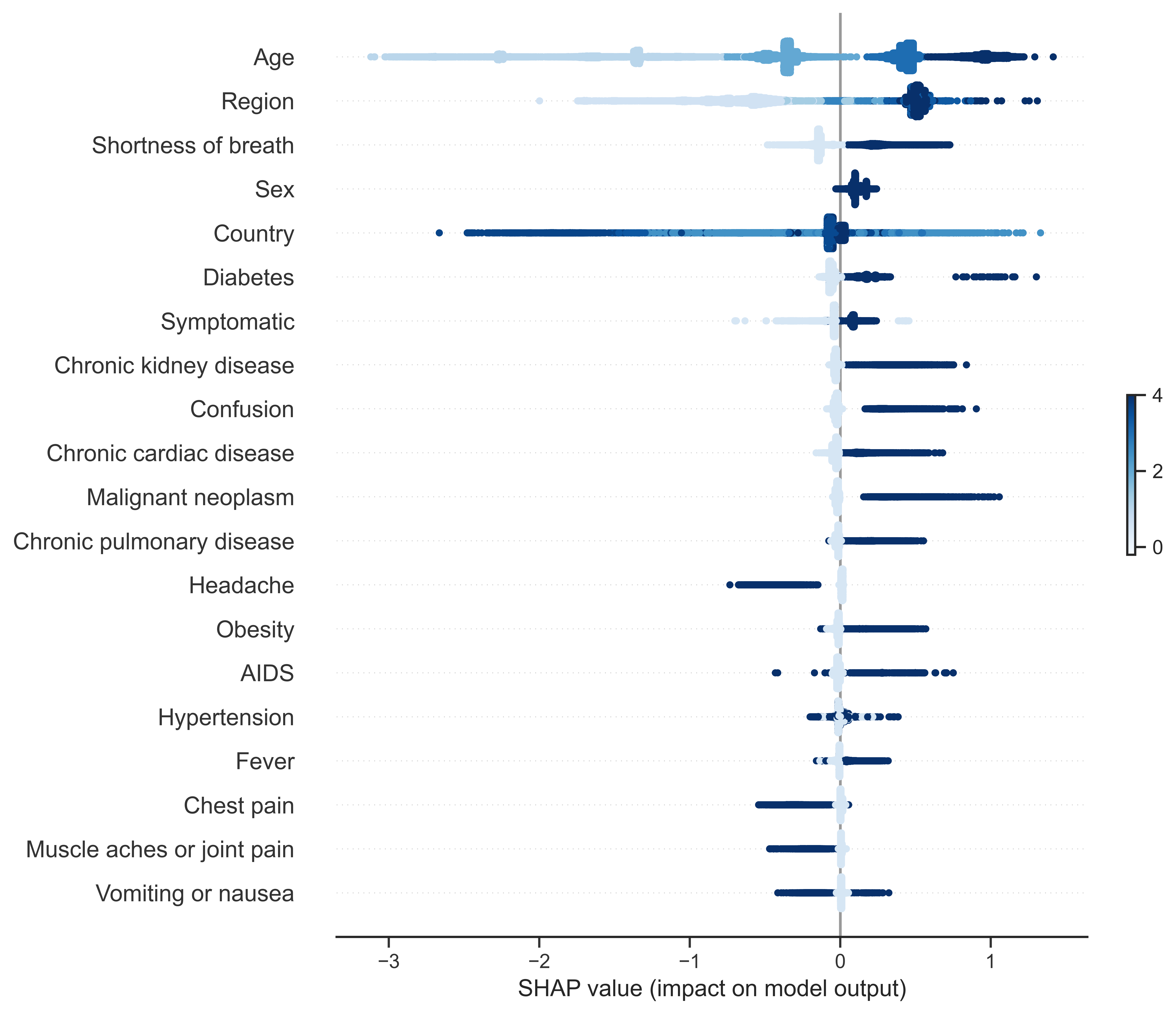}
    \caption{XGBoost Feature Importance with SHAP for Mortality (only Men). The values in the legend being higher or darker colour in the plot correspond to higher values of that feature contributing to the prediction either for stronger positive prediction (more colour points for the feature on the right side of the vertical line) or stronger negative prediction of outcome otherwise.}
    \label{fig:Death_Shap_Men}
\end{figure}

\begin{figure}[h]
    \centering
    \includegraphics[width=1.0\linewidth]{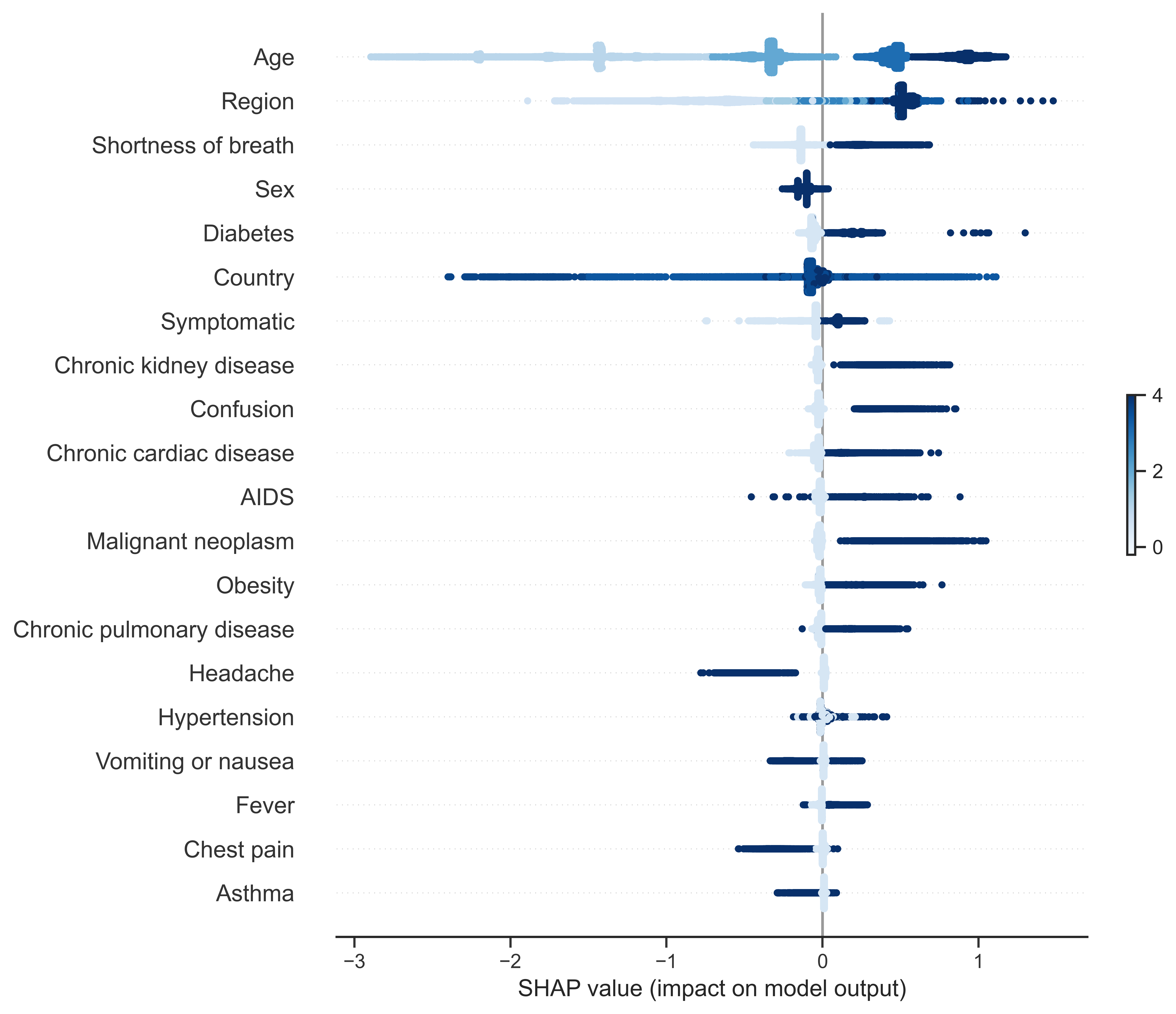}
    \caption{XGBoost Feature Importance with SHAP for Mortality (only Women). The values in the legend being higher or darker colour in the plot correspond to higher values of that feature contributing to the prediction either for stronger positive prediction (more colour points for the feature on the right side of the vertical line) or stronger negative prediction of outcome otherwise.}
    \label{fig:Death_Shap_Women}
\end{figure}

\noindent Taking the best performing XGBoost model and applying 2 different feature importance methods, average f1-score gain across splits and Shapley values, we obtain the results seen in Figures \ref{fig:PE_FI}, \ref{fig:PE_Shap}, \ref{fig:Death_FI}, and \ref{fig:Death_Shap}. A feature importance stratification on a held-out test set of only men and only women separately for either PE or mortality prediction is also included in Figures \ref{fig:PE_Shap_Men}, \ref{fig:PE_Shap_Women}, \ref{fig:Death_Shap_Men}, and \ref{fig:Death_Shap_Women}. As further clarification for the SHAP plot, darker colour indicates that a higher value of that feature contributes to the prediction either positively (if on the right hand side of the vertical line) or negatively (if on the left hand side of the vertical line). Higher placement of the feature vertically in the plot means it has a higher mean Shapley value and hence contributes more to correct predictions in the model. \\

\section{Discussion}
\noindent To our knowledge, this multi-center dataset is the largest international cohort of hospitalised COVID-19 patients available. Our analysis showed that patients with PE are older, more often male, white, from higher income countries, and are more likely to suffer from: asthma, chronic cardiac disease, chronic kidney disease, chronic neurological disease, chronic pulmonary disease, hypertension, cancer, obesity, rheumatologic conditions, and smoking. \\

\noindent The occurrence of pulmonary embolism in our study population was 0.7\% and our results showed a significant association between confirmed PE and mortality when compared with patients without PE as has been similarly found in patients without COVID-19 \cite{gomez2021mortality}. \\

\noindent Accordingly, our logistic regression models for PE and death showed that different age-groups experience different risks of either outcome. The age group 40-80 was at highest odds of having PE, and those $>$60 of dying as can be seen in the Kaplan-Meier curve in Figure \ref{fig:KM_Age}. Symptomatic COVID-19 patients were almost 3 times more likely to experience PE while also being more likely to die. Within symptoms and comorbidities, shortness of breath, chest pain, obesity, and bleeding were associated with higher odds of a PE, followed by hypertension and loss of smell. The regionality of the data must be addressed in the higher odds of death in South Asia, Middle East and North Africa (MENA), and South Africa compared to Europe and Central Asia as the hospital centers in those communities have different challenges and circumstances when it comes to fighting the pandemic. Symptoms like shortness of breath, confusion, severe dehydration, and wheezing were present in COVID-19 patients with higher odds of death, and comorbidities such as malignant neoplasm, diabetes, and chronic kidney or liver disease also lead to higher risk of death. For both correlation and odds of PE and death, men were more at risk. This is shown in the Kaplan-Meier curves for survival stratified across sex in Figure \ref{fig:KM_Sex}. \\

\noindent The hazard ratios confirmed those over the age of 60 were at highest risk of death, especially those COVID-19 patients who experienced shortness of breath, severe dehydration, confusion, and had pre-existing chronic conditions. Regionality of hospital admission was once again an important risk factor for death. Interestingly, patients with PE, chest pain, asthma, or fever seemed to have lower risk associated which could be due to earlier and easier detection of these symptoms and conditions in the progression of the disease. \\

\noindent Seeking to combine this clinically insightful information for outcome prediction, we developed a fast prediction model with XGBoost for both PE and death in COVID-19 hospitalised patients, and tested it in different countries separately. We also showed that appropriate class weighting can help with class imbalance and even outperform ensemble resampling methods without having to sacrifice the interpretability of the model (Tables \ref{tab:PE_UKSpain}, \ref{tab:Death_All}). The best performing model for PE prediction evaluated across separate held-out UK-only data, Spain-only data, and UK and Spain data combined is XGBoost without undersampling and without rigid thresholding using robust class weighting. As for death, the XGBoost again outperformed all other models including the ensemble with XGBoost on some metrics. \\

\noindent Since our XGBoost model outperformed other methods, we also showed that the best method for handling class imbalance is through robust class weighting and compared it to other methods for imbalance handling like ensembles and resampling methods. Another advantage of this method is that it avoids introducing bias like in the case of resampling. Finally, XGBoost provides feature importances which was useful for explaining clinical risk prediction of the model to healthcare professionals and policy-makers. \\

\noindent Exploring two different interpretability methods for XGBoost, average gain across splits and Shapley values, showed that the time of dominant presence of the alpha variant, age, fever, shortness of breath, and hypertension were the key predictors for PE, followed by region of admission, sex, and chest pain. Recent work has alluded to an association between the alpha variant and occurrence of thromboembolisms in mice but further research relevant to human samples is missing \cite{law2022incidence}. Age was a complex non-linear predictor with different age groups corresponding to varying risks. Tthe clear colour separation for the Shapley values for age in Figure \ref{fig:Death_Shap} showed how each age group has a clearly separable predictive value for mortality with older groups having higher risk but which is not the case for PE as younger age groups can be more predictive of higher PE risk. Furthermore, Shapley values analysis identified obesity, smoking, and the presence of cough as important predictors for PE whereas the default XGBoost method does not. The most predictive features for all-cause mortality were age, region of hospital admission, sex, diabetes, and shortness of breath whereas the default method highlights hypertension and obesity in addition. For mortality, higher values of region corresponded to samples from South Asia and South Africa. \\

\begin{table*}[t]
    \small
    \centering
    \caption{\textbf{Features of Significant Importance for PE and Mortality Prediction According to Different Models (For XGBoost Top 20 SHAP Value Features Were Taken as Important, and for Logistic Regression and Cox model significance was taken as p$<$0.005). Ticks correspond to significance of feature for that model and for that outcome and X corresponds to lack of significance. Features in bold are those found to be significant for both mortality and PE prediction.}}
    \begin{tabularx}{0.6\linewidth}{l
                                    c
                                    c
                                    c
                                    c
                                    c
                                    c
                                    c
                                 }
         \toprule
        & \multicolumn{2}{c}{PE} && \multicolumn{3}{c}{Death}  \\
        \cmidrule(r){2-3} \cmidrule(l){4-7}
         \bf{Feature}  &  \multicolumn{1}{c}{\bf{LR}} & \bf{XGBoost} &&  \multicolumn{1}{c}{\bf{\bf{LR}}} & \bf{Cox} & \bf{XGBoost}\\
            \midrule
          \textbf{Age} & \checkmark & \checkmark &&  \checkmark & \checkmark & \checkmark\\
          \addlinespace[0.05cm]
          \textbf{Sex} & \checkmark & \checkmark && \checkmark & \checkmark & \checkmark\\
          \addlinespace[0.05cm]
          \textbf{Region} & \checkmark & \checkmark && \checkmark & \checkmark & \checkmark\\
          \addlinespace[0.05cm]
          \textbf{Alpha Variant} & \checkmark & \checkmark &&  &  & \\
          \addlinespace[0.05cm]
          \textit{Comorbidities} &  &  &&  & &\\
          \addlinespace[0.05cm]
          \hspace{0.5cm} AIDS/HIV & X & X && \checkmark & \checkmark & \checkmark\\
          \addlinespace[0.05cm]
          \hspace{0.5cm} Asthma & X & X && \checkmark & \checkmark & X\\
          \addlinespace[0.05cm]
          \hspace{0.5cm} \textbf{Chronic Cardiac Disease} & \checkmark & \checkmark && \checkmark & \checkmark & \checkmark\\
          \addlinespace[0.05cm]
          \hspace{0.5cm} Chronic Haematological & X & X && \checkmark & X & X\\
          \addlinespace[0.05cm]
          \hspace{0.5cm} \textbf{Chronic Kidney Disease} & \checkmark & \checkmark && \checkmark & \checkmark & \checkmark\\
          \addlinespace[0.05cm]
          \hspace{0.5cm} Chronic Neurological & X & X && \checkmark & \checkmark & X\\
          \addlinespace[0.05cm]
          \hspace{0.5cm} \textbf{Chronic Pulmonary} & \checkmark & \checkmark && \checkmark & \checkmark & \checkmark\\
          \addlinespace[0.05cm]
          \hspace{0.5cm} \textbf{Dementia} & \checkmark & \checkmark && \checkmark & \checkmark & X\\
          \addlinespace[0.05cm]
          \hspace{0.5cm} \textbf{Diabetes} & \checkmark & \checkmark && \checkmark & \checkmark & \checkmark\\
          \addlinespace[0.05cm]
          \hspace{0.5cm} \textbf{Hypertension} & \checkmark & \checkmark && \checkmark & X & \checkmark\\
          \addlinespace[0.05cm]
          \hspace{0.5cm} Liver Disease & X & X && \checkmark & \checkmark & X\\
          \addlinespace[0.05cm]
          \hspace{0.5cm} \textbf{Malignant Neoplasm} & \checkmark & \checkmark && \checkmark & \checkmark & \checkmark\\
          \addlinespace[0.05cm]
          \hspace{0.5cm} Malnutrition & \checkmark & X && \checkmark & \checkmark & X\\
          \addlinespace[0.05cm]
          \hspace{0.5cm} \textbf{Obesity} & \checkmark & \checkmark && \checkmark & \checkmark & \checkmark\\
          \addlinespace[0.05cm]
          \hspace{0.5cm} Rheumatologic & X & X && \checkmark & \checkmark & X\\
          \addlinespace[0.05cm]
          \hspace{0.5cm} \textbf{Smoking} & \checkmark & \checkmark && X & X & X\\
            \bottomrule
    \end{tabularx}
    \label{tab:Importance1}
\end{table*}

\begin{table*}[t]
    \small
    \centering
    \caption{\textbf{Features of Significant Importance for PE and Mortality Prediction According to Different Models (For XGBoost Top 20 SHAP Value Features Were Taken as Important, and for Logistic Regression and Cox model significance was taken as p$<$0.005). Ticks correspond to significance of feature for that model and for that outcome and X corresponds to lack of significance. Features in bold are those found to be significant for both mortality and PE prediction. (continued)}}
    \begin{tabularx}{0.6\linewidth}{@{}l
                                    c
                                    c
                                    c
                                    c
                                    c
                                    c
                                    c
                                 @{}}
         \toprule
        & \multicolumn{2}{c}{PE} && \multicolumn{3}{c}{Death}  \\
        \cmidrule(r){2-3} \cmidrule(l){4-7}
         \bf{Feature}  &  \multicolumn{1}{c}{\bf{LR}} & \bf{XGBoost} &&  \multicolumn{1}{c}{\bf{\bf{LR}}} & \bf{Cox} & \bf{XGBoost}\\
            \midrule
          \textit{Symptoms} &  &  &&  & &\\
          \addlinespace[0.05cm]
          \hspace{0.5cm} \textbf{Symptomatic} & \checkmark & \checkmark && \checkmark & \checkmark & \checkmark\\
          \addlinespace[0.05cm]
          \hspace{0.5cm} Abdominal Pain & \checkmark & X && \checkmark & \checkmark & X\\
          \addlinespace[0.05cm]
          \hspace{0.5cm} Confusion & \checkmark & X && \checkmark & \checkmark & \checkmark\\
          \addlinespace[0.05cm]
          \hspace{0.5cm} Bleeding & \checkmark & X && \checkmark & \checkmark & X\\
          \addlinespace[0.05cm]
          \hspace{0.5cm} \textbf{Chest Pain} & \checkmark & \checkmark && \checkmark & \checkmark & \checkmark\\
          \addlinespace[0.05cm]
          \hspace{0.5cm} Conjunctivitis & X & X && \checkmark & \checkmark & X\\
          \addlinespace[0.05cm]
          \hspace{0.5cm} Cough & X & \checkmark && \checkmark & \checkmark & X\\
          \addlinespace[0.05cm]
          \hspace{0.5cm} Diarrhoea & X & X && \checkmark & \checkmark & X\\
          \addlinespace[0.05cm]
          \hspace{0.5cm} Ear Pain & X & X && X & X & X\\
          \addlinespace[0.05cm]
          \hspace{0.5cm} \textbf{Fatigue} & \checkmark & \checkmark && \checkmark & \checkmark & X\\
          \addlinespace[0.05cm]
          \hspace{0.5cm} \textbf{Headache} & \checkmark & \checkmark && \checkmark & \checkmark & \checkmark\\
          \addlinespace[0.05cm]
          \hspace{0.5cm} \textbf{Fever} & \checkmark & \checkmark && \checkmark & \checkmark & \checkmark\\
          \addlinespace[0.05cm]
          \hspace{0.5cm} Lost Sense of Smell & \checkmark & X && \checkmark & \checkmark & X\\
          \addlinespace[0.05cm]
          \hspace{0.5cm} Lost Sense of Taste & X & X && \checkmark & \checkmark & X\\
          \addlinespace[0.05cm]
          \hspace{0.5cm} Lymphadenopathy & X & X && \checkmark & \checkmark & X\\
          \addlinespace[0.05cm]
          \hspace{0.5cm} \textbf{Muscle/Joint Pain} & \checkmark & X && \checkmark & \checkmark & \checkmark\\
          \addlinespace[0.05cm]
          \hspace{0.5cm} Runny Nose & X & X && \checkmark & X & X\\
          \addlinespace[0.05cm]
          \hspace{0.5cm} Seizures & \checkmark & X && X & X & X\\
          \addlinespace[0.05cm]
          \hspace{0.5cm} Severe Dehydration & \checkmark & X && \checkmark & \checkmark & X\\
          \addlinespace[0.05cm]
          \hspace{0.5cm} \textbf{Shortness of Breath} & \checkmark & \checkmark && \checkmark & \checkmark & \checkmark\\
          \addlinespace[0.05cm]
          \hspace{0.5cm} Skin Rash & X & X && \checkmark & \checkmark & X\\
          \addlinespace[0.05cm]
          \hspace{0.5cm} Sore Throat & \checkmark & X && X & X & X\\
          \addlinespace[0.05cm]
          \hspace{0.5cm} \textbf{Vomiting} & \checkmark & \checkmark && \checkmark & \checkmark & \checkmark\\
          \addlinespace[0.05cm]
          \hspace{0.5cm} Wheezing & \checkmark & \checkmark && \checkmark & \checkmark & X\\
          \addlinespace[0.05cm]
          PE & - & - && \checkmark & \checkmark & X\\
            \bottomrule
    \end{tabularx}
    \label{tab:Importance2}
\end{table*}

\noindent The pulmonary embolism and mortality prediction model can help with management of COVID-19 as it uses standard demographics, comorbidity, and symptom data collected at admission for identifying patients most at risk of developing PE which may enable an earlier start of targeted anticoagulation therapy. Our mortality risk prediction model can also help with patient population risk assessment and prioritisation across different regions of the world. \\

\noindent A strength of the current study is that a combination of machine learning and traditional statistical modeling can offer a more reliable system for predictive risk forecasting. XGBoost provides at-admission prediction of both events, while odds and hazards ratios obtained from logistic regression and the Cox proportional hazards model give us an insight into stratified risk and global feature importance. We systematically compare our XGBoost model with different risk prediction algorithms. Our model also outperforms recently published results across a variety of metrics like AUROC and sensitivity despite being developed on a much larger and more heterogeneous and diverse dataset while being robust to class imbalance \cite{ikemura2021using}. With existing scores built on non-COVID-19 data like The National Early Warning Score 2, there is insufficient information available on their reliability in the COVID-19 setting, and some have been found to underestimate mortality \cite{alballa2021machine}. Our model is able to deploy at admission for both PE and death risk prediction and can help supplant these needs rapidly. \\

\noindent The study, however, has several limitations. First, almost 60\% of patients who died did so in South Africa, and over 70\% of PE cases were located in the UK. This may be due to limited access to d-dimer tests or CT scans. There were no mandatory diagnostic criteria in the ISARIC CRF for PE. The absence of a control group of patients without COVID-19 in this dataset prevented estimation of specificity. The patient cohort comprised of hospitalised patients with confirmed COVID-19 who had a mortality rate of 21.7\%. These models are not for use in the community and could still perform differently in populations at lower risk of death and across different regions of the world. \\

\noindent In conclusion, the set of decisions taken after must include different stakeholders like patients, clinicians, hospital administrators, researchers, and data procurers so that trade-offs can be identified and context-informed decisions can be taken to address them, especially if our models could have missed harms or benefits to different groups and communities. \\


%

\section{Contributors}
\noindent CK conceived and designed the study. The data was curated by LM and BC. Formal analysis was undertaken by MM. Development of the statistical analysis and machine learning methodologies was completed by MM, LC, and CK. Project administration was done by CK, LC, LM, and BC. Software was developed by MM, and validated by MM and CK. Supervision was provided by CK and LC. Visualisations, writing, and editing was done by MM. Resources, clinical or otherwise, was provided by LM, PO, XW, GR, KP, and FG. All authors subsequently critically edited the report. All authors read and approved the final report. The corresponding author and CK had full access to all data. MM and CK accessed and verified the data and results. MM and CK had final responsibility for the decision to submit for publication. \\

\section{Declaration of interests}
\noindent GR declares receiving a grant from United States National Institute of Health, R01 Grant: Emerging Zoonotic Malaria in Malaysia: Strenghtening Surveillance and Evaluating Population Genetics Structure to Improve Regional Risk Prediction Tool and travel support from the European Society of Clinical Microbiology and Infectious Disease (ESCMID) for observership at European Centre for Disease Prevention and Control (ECDC). All authors declare no competing interests. \\

\section{Data sharing}
\noindent The data that underpin this analysis are highly detailed clinical data on individuals hospitalised with COVID-19. Due to the sensitive nature of these data and the associated privacy concerns, they are available via a governed data access mechanism following review of a data access committee. Data can be requested via the IDDO COVID-19 Data Sharing Platform (http://www.iddo.org/covid-19). The Data Access Application, Terms of Access and details of the Data Access Committee are available on the website. Briefly, the requirements for access are a request from a qualified researcher working with a legal entity who have a health and/or research remit; a scientifically valid reason for data access which adheres to appropriate ethical principles. The full terms are at https://www.iddo.org/document/covid-19-data-access-guidelines. A small subset of sites who contributed data to this analysis have not agreed to pooled data sharing as above. In the case of requiring access to these data, please contact the corresponding author in the first instance who will look to facilitate access. 

\section{Acknowledgements}
\noindent M. Mesinovic appreciates the support of the EPSRC Center for Doctoral Training in Health Data Science (EP/S02428X/1) and the Rhodes Trust \\
\noindent This work was made possible by the UK Foreign, Commonwealth and Development Office and Wellcome [215091/Z/18/Z, 222410/Z/21/Z, 225288/Z/22/Z and 220757/Z/20/Z]; the Bill \& Melinda Gates Foundation [OPP1209135]; the philanthropic support of the donors to the University of Oxford’s COVID-19 Research Response Fund (0009109); CIHR Coronavirus Rapid Research Funding Opportunity OV2170359 and the coordination in Canada by Sunnybrook Research Institute; endorsement of the Irish Critical Care-Clinical Trials Group, co-ordination in Ireland by the Irish Critical Care-Clinical Trials Network at University College Dublin and funding by the Health Research Board of Ireland [CTN-2014-12]; the Rapid European COVID-19 Emergency Response research (RECOVER) [H2020 project 101003589] and European Clinical Research Alliance on Infectious Diseases (ECRAID) [965313]; the COVID clinical management team, AIIMS, Rishikesh, India; the COVID-19 Clinical Management team, Manipal Hospital Whitefield, Bengaluru, India; Cambridge NIHR Biomedical Research Centre; the dedication and hard work of the Groote Schuur Hospital Covid ICU Team and supported by the Groote Schuur nursing and University of Cape Town registrar bodies coordinated by the Division of Critical Care at the University of Cape Town; the Liverpool School of Tropical Medicine and the University of Oxford; the dedication and hard work of the Norwegian SARS-CoV-2 study team and the Research Council of Norway grant no 312780, and a philanthropic donation from Vivaldi Invest A/S owned by Jon Stephenson von Tetzchner; Imperial NIHR Biomedical Research Centre; the Comprehensive Local Research Networks (CLRNs) of which PJMO is an NIHR Senior Investigator (NIHR201385); Innovative Medicines Initiative Joint Undertaking under Grant Agreement No. 115523 COMBACTE, resources of which are composed of financial contribution from the European Union’s Seventh Framework Programme (FP7/2007- 2013) and EFPIA companies, in-kind contribution; Stiftungsfonds zur Förderung der Bekämpfung der Tuberkulose und anderer Lungenkrankheiten of the City of Vienna, Project Number: APCOV22BGM; Italian Ministry of Health “Fondi Ricerca corrente–L1P6” to IRCCS Ospedale Sacro Cuore–Don Calabria; Australian Department of Health grant (3273191); Gender Equity Strategic Fund at University of Queensland, Artificial Intelligence for Pandemics (A14PAN) at University of Queensland, the Australian Research Council Centre of Excellence for Engineered Quantum Systems (EQUS, CE170100009), the Prince Charles Hospital Foundation, Australia; grants from Instituto de Salud Carlos III, Ministerio de Ciencia, Spain; Brazil, National Council for Scientific and Technological Development Scholarship number 303953/2018-7; the Firland Foundation, Shoreline, Washington, USA;  the French COVID cohort (NCT04262921) is sponsored by INSERM and is funded by the REACTing (REsearch \& ACtion emergING infectious diseases) consortium and by a grant of the French Ministry of Health (PHRC n20-0424); a grant from foundation Bevordering Onderzoek Franciscus; the South Eastern Norway Health Authority and the Research Council of Norway; Institute for Clinical Research (ICR), National Institutes of Health (NIH) supported by the Ministry of Health Malaysia; preparedness work conducted by the Short Period Incidence Study of Severe Acute Respiratory Infection; the U.S. DoD Armed Forces Health Surveillance Division, Global Emerging Infectious Diseases Branch to the U.S Naval Medical Research Unit No. TWO (NAMRU-2) (Work Unit \#: P0153\_21\_N2). These authors would like to thank Vysnova Partners, Inc. for the management of this research project. The Lao-Oxford-Mahosot Hospital-Wellcome Trust Research Unit is funded by the Wellcome Trust. \\
\noindent This work uses data provided by patients and collected by the NHS as part of their care and support \#DataSavesLives. The data used for this research were obtained from ISARIC4C. We are extremely grateful to the 2648 frontline NHS clinical and research staff and volunteer medical students who collected these data in challenging circumstances; and the generosity of the patients and their families for their individual contributions in these difficult times. The COVID-19 Clinical Information Network (CO-CIN) data was collated by ISARIC4C Investigators. Data and Material provision was supported by grants from: the National Institute for Health Research (NIHR; award CO-CIN-01), the Medical Research Council (MRC; grant MC\_PC\_19059), and by the NIHR Health Protection Research Unit (HPRU) in Emerging and Zoonotic Infections at University of Liverpool in partnership with Public Health England (PHE), (award 200907), NIHR HPRU in Respiratory Infections at Imperial College London with PHE (award 200927), Liverpool Experimental Cancer Medicine Centre (grant C18616/A25153), NIHR Biomedical Research Centre at Imperial College London (award ISBRC-1215-20013), and NIHR Clinical Research Network providing infrastructure support. We also acknowledge the support of Jeremy J Farrar and Nahoko Shindo. \\

\ifCLASSOPTIONcaptionsoff
  \newpage
\fi



%
\bibliographystyle{vancouver} 
\bibliography{References}

%




\end{document}